\documentclass[,
twocolumn,
superscriptaddress,
amsmath,amssymb,
prc,
]{revtex4}
\usepackage{mathrsfs}
\usepackage[pdftex]{graphicx}
\usepackage{dcolumn}
\usepackage{bm}
\usepackage{longtable}
\usepackage{color,hyperref}
\usepackage{longtable}
\hypersetup{colorlinks,breaklinks,linkcolor=blue,urlcolor=blue,anchorcolor=blue,citecolor=blue}
\usepackage{wrapfig}
\usepackage{lscape}
\usepackage{rotating}
\usepackage{multirow}            

\newcolumntype{d}[1]{D{.}{.}{#1}}

\begin{document}
\title{In-beam spectroscopy of medium- and high-spin states in $^{133}$Ce}

\author{A. D. Ayangeakaa}
\altaffiliation[Present address: ]{Physics Division, Argonne National Laboratory, Argonne, Illinois 60439, USA.} 
\author{U. Garg}
\affiliation{Department of Physics, University of Notre Dame, Notre Dame, Indiana 46556, USA}  
\author{C. M. Petrache}
\author{S. Guo}
\altaffiliation[On leave from: ]{Institute of Modern Physics, Chinese Academy of Sciences, Lanzhou 730000, China.} 
\affiliation{Centre de Sciences Nucl\'eaires et Sciences de la Mati\`ere, CNRS/IN2P3, Universit\'{e} Paris-Saclay, B\^at. 104-108, 91405  Orsay, France} 
\author{P. W. Zhao}%
\affiliation{Physics Division, Argonne National Laboratory, Argonne, Illinois 60439, USA}
\author{J. T. Matta}
\altaffiliation[Present address: ]{Physics Division, Oak Ridge National Laboratory, Oak Ridge, Tennessee 37830, USA.}
\affiliation{Department of Physics, University of Notre Dame, Notre Dame, Indiana 46556, USA}  %
\author{B. K. Nayak}
\altaffiliation[Present address: ]{Nuclear Physics Division, Bhabha Atomic Research Center (BARC), Mumbai 400085, India.}
\affiliation{Department of Physics, University of Notre Dame, Notre Dame, Indiana 46556, USA}  %
\author{D. Patel}
\altaffiliation[Present address: ]{M.D. Anderson Cancer Center, Houston, Texas 77030, USA.}
\affiliation{Department of Physics, University of Notre Dame, Notre Dame, Indiana 46556, USA}  %

\author{R. V. F. Janssens}
\author{M. P. Carpenter}
\affiliation{Physics Division, Argonne National Laboratory, Argonne, Illinois 60439, USA}
\author{C. J. Chiara}
\altaffiliation[Present Address: ]{U.S. Army Research Laboratory, Adelphi, Maryland 20783, USA.}
\affiliation{Physics Division, Argonne National Laboratory, Argonne, Illinois 60439, USA}
\affiliation{Department of Chemistry and Biochemistry, University of Maryland, College Park, Maryland 20742, USA}

\author{F. G. Kondev}
\affiliation{Nuclear Engineering Division, Argonne National Laboratory, Argonne, Illinois 60439, USA}
\author{T. Lauritsen}
\author{D. Seweryniak}
\author{S. Zhu}
\affiliation{Physics Division, Argonne National Laboratory, Argonne, Illinois 60439, USA}%
\author{S. S. Ghugre}
\affiliation{UGC-DAE Consortium for Science Research, Kolkata 700098, India}
\author{R. Palit}
\affiliation{Tata Institute of Fundamental Research, Mumbai 400005, India}
\affiliation{The Joint Institute for Nuclear Astrophysics, University of Notre Dame, Notre Dame, Indiana 46556, USA}

\begin{abstract}
Medium and high-spin states in $^{133}$Ce were investigated  using  the  $^{116}$Cd($^{22}$Ne,5$n$) reaction and the Gammasphere array. The level scheme has been extended up to an excitation energy of $\sim$ 22.8 MeV and spin 93/2 $\hbar$. Eleven bands of quadrupole transitions and two new dipole bands are identified. The connections to low-lying states of the previously known, high-spin triaxial bands were firmly established, thus fixing the excitation energy and, in many cases, the spin/parity of the levels. Based on comparisons with Cranked Nilsson-Strutinsky (CNS) calculations and tilted axis cranking covariant density functional theory (TAC-CDFT), it is shown that all observed bands are characterized by pronounced triaxiality. Competing multiquasiparticle configurations are found to contribute to a rich variety of collective phenomena in this nucleus. 
 
  \end{abstract}



\maketitle

\section{Introduction}
The $^{133}$Ce nucleus has been the focus of extensive experimentation and theoretical investigations for a long time and a number of important collective phenomena have been uncovered. Most recently, this nucleus  was studied using the $^{116}$Cd($^{22}$Ne,5$n$) reaction and the Gammasphere array \cite{133ce-a}, leading to the identification of three new dipole bands, which represent the first experimental evidence for the multiple chiral doublet bands (M$\chi$D) phenomenon.  Prior works  on $^{133}$Ce reported results mainly on the medium-spin states in this nucleus. In the earliest experiment, seven bands were identified and the level scheme extended up to spin 49/2 \cite{ma1987}. The observed rotational bands were discussed in the framework of the cranking model and configurations based on one- and three-quasiparticle excitations were assigned. One sequence with quadrupole transitions only was observed, but not linked to low-lying states. Its three-quasiparticle configuration was suggested to involve either the $ \nu i_{13/2}$ or the $ \nu f_{7/2}$ orbital. Subsequently, the lifetimes of the states of the $ \nu h_{11/2}$ yrast band and one of the three quasiparticle bands were measured and the results confirmed the previously proposed configuration assignments~\cite{emediato1997}. 

The first study of the high-spin level structure of  $^{133}$Ce was performed using the Gammasphere array, and revealed the existence of three superdeformed bands \cite{karl1995}.  The interpretation in the cranking approximation suggested superdeformed configurations involving one $\nu i_{13/2}$ or $ \nu f_{7/2}$ neutron coupled to the $^{132}$Ce  superdeformed core. In addition, six new rotational structures were identified at high spins, with characteristics of triaxial configurations \cite{karl1996}. However, none of these bands were linked to low-lying states. The measured lifetimes of one of these bands permitted the extraction of a transitional quadrupole moment of 2.2 $\it e$b, thus confirming the triaxial interpretation~\cite{joss1998}.

The present paper reports on new experimental results that relate to both high- and low-spin structures in $^{133}$Ce.  First, the high-spin triaxial bands reported in Ref.~\cite{karl1996} are now firmly connected to low-lying states through the identification of several linking transitions, thereby establishing their excitation energy, spin and parity. However, many transitions of the previously reported triaxial sequences are now placed differently. The level scheme is also extended to a higher excitation energy and spin of 22.8 MeV and $93/2$ $\hbar$, respectively. Secondly, two dipole bands and four rotational sequences  of $\Delta I=2$ transitions are newly identified, and the angular-distribution coefficients and anisotropies of several transitions have been determined. The observed collective structures are extensively discussed in the framework of the Cranked Nilsson-Strutinsky (CNS) model, as described in Refs.~\cite{ing-phys-rep,Ben85,Afa95,Car06} and, one band of dipole character is interpreted using the tilted axis cranking covariant density functional theory (TAC-CDFT)~\cite{Zhao2011Phys.Lett.B181,Meng2013FrontiersofPhysics55}. A consistent interpretation of most of the observed bands is achieved. The observed level structure of  $^{133}$Ce illustrates the ability of nuclei in the $A=130$ mass region to acquire different shapes and to rotate around either the principal or tilted axes of the intrinsic frame, as is the case in the neighboring  $^{138-141}$Nd nuclei for which new results were recently reported~\cite{138-low,138-switch,138-high,140-high,141nd}.

\section{\label{sec-exp} Experimental details}

The present work documents new and extended results at medium and high spins and continues the study of the $^{133}$Ce nucleus, a salient feature of which, the observation of multiple chiral bands, was published previously~\cite{133ce-a}. Both studies are based on the same measurement and hence, the experimental procedure and the analysis methods are similar, but with more information provided here. 

The experiment was performed at the ATLAS facility at Argonne National Laboratory. Medium- and high-spin states in $^{133}$Ce were populated in two separate experiments following the $^{116}$Cd($^{22}$Ne,5$n$) reaction. In the first, a 112-MeV beam of $^{22}$Ne bombarded a 1.48 mg/cm$^{2}$-thick  target foil of isotopically enriched $^{116}$Cd, sandwiched between a 50 $\mathrm{\mu g/cm^{2}}$ thick front layer of Al and a 150 $\mathrm{\mu g/cm^{2}}$ Au backing. The second experiment used the same beam and a target of the same enrichment and thickness evaporated onto a  55 $\mathrm{\mu g/cm^{2}}$-thick Au foil. A combined total of  $4.1\times10^{9}$ four- and higher-fold prompt $\gamma$-ray  coincidence events were accumulated using the Gammasphere array~\cite{LEE1990c641}, which comprised 101 (88) active Compton-suppressed HPGe detectors during the first (second) experiment. The accumulated events were unfolded and sorted into fully symmetrized, three-dimensional ($E_{\gamma}$-$E_{\gamma}$-$E_{\gamma}$) and four-dimensional ($E_{\gamma}$-$E_{\gamma}$-$E_{\gamma}$-$E_{\gamma}$) histograms and analyzed using the \textsc{radware}~\cite{rad1,rad2} analysis package. 

Multipolarity assignments were made on the basis of extensive angular-distribution measurements~\cite{Iacob199757} and, for weak transitions, on a two-point angular-correlation ratio, $R_{ac}$~\cite{KramerFlecken1989333,Chiara.75.054305}. The angular-distribution analysis was performed using coincidence matrices sorted such that energies of $\gamma$ rays detected at specific Gammasphere angles (measured with respect to the beam direction) $E_{\gamma}(\theta)$, were incremented on one axis, while the energies of coincident transitions detected at any angle, $E_{\gamma}(any)$, were placed on the other. To improve statistics, adjacent rings of Gammasphere and those corresponding to angles symmetric with respect to $90^\circ$ in the forward and backward hemispheres were combined. A total of seven matrices (with the angles 17.3$^\circ$, 34.6$^\circ$, 50.1$^\circ$, 58.3$^\circ$, 69.8$^\circ$, 80.0$^\circ$, and 90.0$^\circ$) were created. After gating on the $E_{\gamma}(any)$ axis, background-subtracted and efficiency-corrected spectra were generated. From these, the intensities of transitions of interest were extracted and fitted to the usual angular distribution function $W(\theta)=a_{o}[1+a_2P_2(cos\theta) + a_4P_4(cos\theta)]$, where $P_2$ and $P_4$ are Legendre polynomials. The extracted coefficients, $a_2$ and $a_4$, contain information on the transition multipolarity. 

The two-point angular correlation ratio, $R_{ac}$ was deduced from a normalized ratio of $\gamma$-ray intensities observed in detectors in the forward or backward angles to the intensities in those centered around $90^\circ$. For this purpose, two coincident matrices were incremented: In the first, $E_{\gamma}(f/b)$-vs-$E_{\gamma}(any)$, detectors in the forward and backward angles were combined and the matrix incremented such that $\gamma$ rays detected at the 31.7$^\circ$, 37.4$^\circ$, 142.6$^\circ$, 148.3$^\circ$, and 162.7$^\circ$ angles were placed on one axis, with transitions observed at any angle grouped along the other. The second matrix, $E_{\gamma}(\sim90^\circ)$-vs-$E_{\gamma}(any)$, was incremented in a similar fashion, but with transitions observed in detectors at 79.2$^\circ$, 80.7$^\circ$, 90.0$^\circ$, 99.3$^\circ$, and 100.8$^\circ$ degrees placed on one axis. The two-dimensional angular correlation ratio, defined by $R_{ac}$ = $I_{\gamma} (\theta_{f/b}, any)$/$I_{\gamma}(\theta_{ \sim90^\circ},any)$, where $I_{\gamma}(\theta_x, any)$ is the $\gamma$-ray intensity obtained by placing gates on the corresponding $E_{\gamma}(any)$ axis. The  $R_{ac}$ ratio, which is independent of the multipolarity of the gating transition and calibrated with transitions of known multipolarity, was established to be greater than 1.0 for stretched-quadrupole and less than $0.8$ for stretched-dipole transitions. \begin{sidewaysfigure*}[]
\vspace{8cm}
\includegraphics[width=\textwidth]{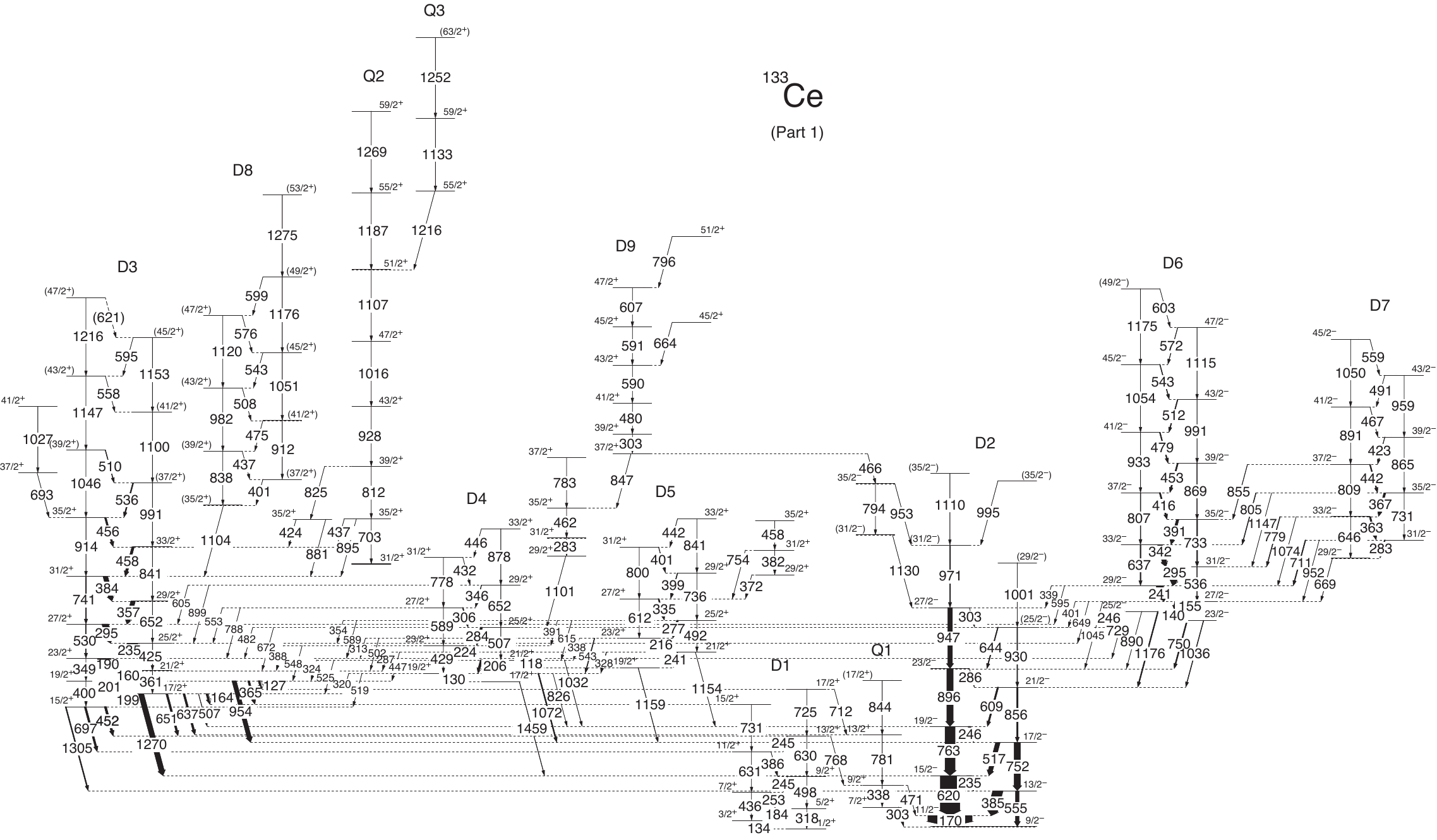}
\caption {\label{FIG1} Part 1 of the level scheme of  $^{133}$Ce. The widths of the arrows are proportional to the relative intensities of the $\gamma$ rays.}
\end{sidewaysfigure*}

\section{\label{sec-res} Results and level scheme}
\begin{figure*}[]
\includegraphics[width=\textwidth]{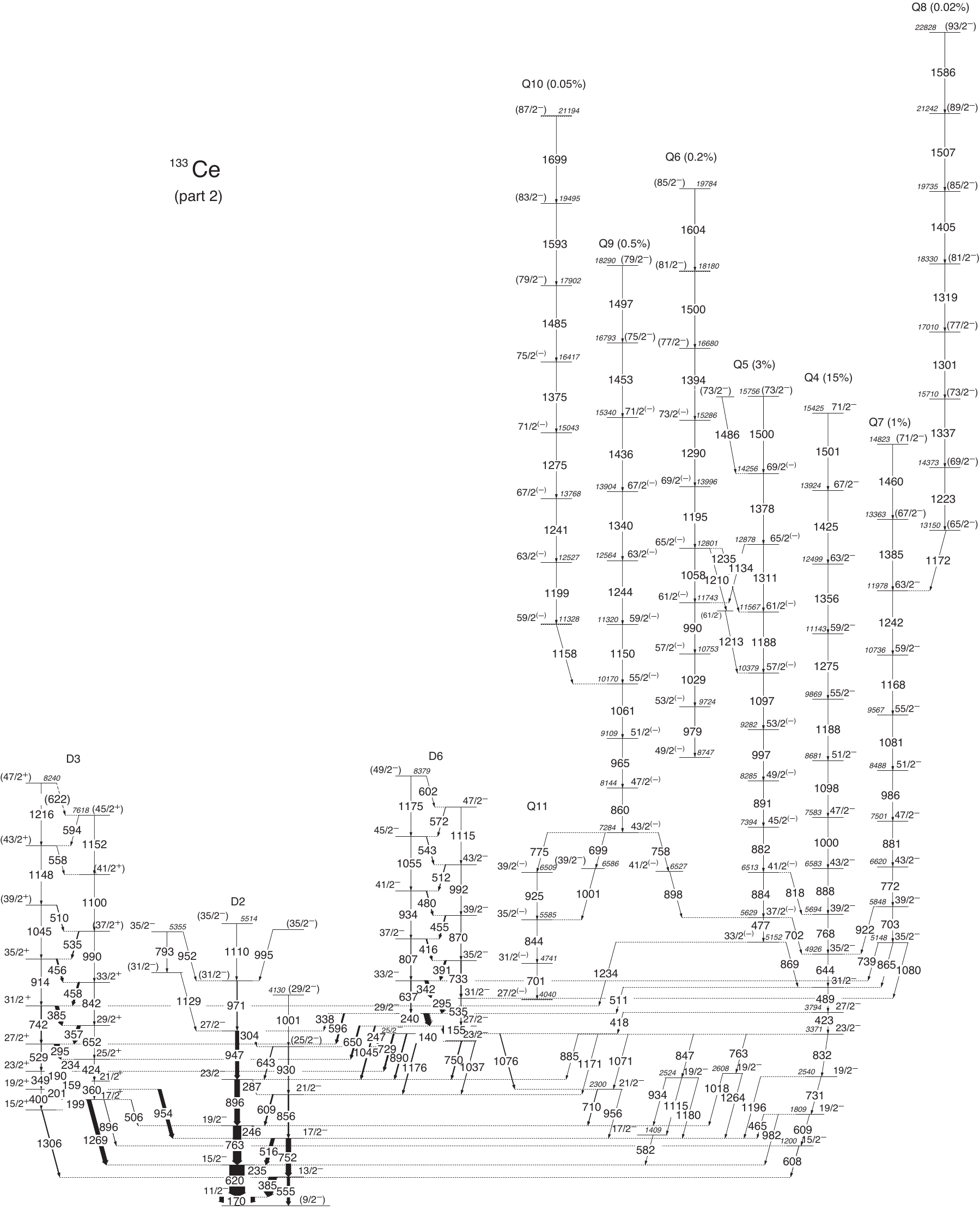}
\caption {\label{FIG2} Part 2 of the level scheme of  $^{133}$Ce. The widths of the arrows are proportional to the relative intensities of the $\gamma$ rays. The intensities of the $Q$ bands are given as a percentage relative to that of the yrast band $D2$.  }
\end{figure*}

The level scheme for $^{133}$Ce, deduced in the present work, builds substantially upon the structure  reported previously in Refs.~\cite{ma1987,karl1995,karl1996}. A portion of the full level scheme, showing the two new dipole bands ($D8$ and $D9$) and three of the four new quadrupole sequences ($Q1$, $Q2$ and $Q3$), is presented in Fig.~\ref{FIG1}. Figure~\ref{FIG2} displays the previously-known, high-spin bands ($Q4$-$Q10$), the newly observed quadrupole sequence $Q11$ and the low-lying states populated by these structures. Transition and level energies, angular-distribution coefficients and anisotropies, as well as the proposed spin and parity of the levels linked by the $\gamma$ rays of interest are summarized in Table \ref{table1}. Since most of the low-spin scheme is already well known and discussed extensively, in the following, the focus is on the newly observed bands and their decay to the low-spin levels. It should also be noted that, while the intensities of all transitions are not reported, those of the $Q$ bands were estimated relative to the yrast band ($D2$) and are indicated in parentheses beside the band labels in Fig.~\ref{FIG2}. The uncertainty in the transition energies is $\sim$0.2 keV for transitions below 1000 keV and intensities larger than 5\% of the $^{133}$Ce reaction channel, $\sim$0.5 keV for those above 1000 keV with intensities lower than 5\%, and $\sim$1 keV for those above 1200 keV and/or weaker than 1\%. Tentative spins and parities assigned to the weakly populated states  are based on  their theoretical interpretation. 
 \\ 

\renewcommand{\thefootnote}{\alph{footnote}}
\begin{center}
\begin{longtable*}{@{\extracolsep{\fill}} rrrrrclc @{\extracolsep{\fill}}}
\caption[] {$\gamma$-ray energies, energies of initial level, angular-distribution coefficients, anisotropies, multipolarities and spin-parity assignments of $\gamma$-ray transitions in $^{133}$Ce. Transitions marked with asterisk signs are unresolved doublets in the multipolarity measurement. 
}\label{table1}\\
\hline \hline

\multicolumn{1}{r}{ Band }&\multicolumn{1}{c}{$E_{\gamma}$ (keV)} & \multicolumn{1}{c}{$ E_i $ (keV)}  & \multicolumn{1}{c}{$ I^{\pi}_i  \rightarrow  I^{\pi}_f$} &  \multicolumn{1}{c}{$a_2$} & \multicolumn{1}{c}{$a_4$} & \multicolumn{1}{c}{ $R_{ac}$} & \multicolumn{1}{c}{Mult.}\\ \hline 
\endfirsthead

\hline \hline
\multicolumn{1}{r}{ Band } &\multicolumn{1}{c}{$E_{\gamma}$ (keV)}  & \multicolumn{1}{c}{$ E_i $ (keV)}  & \multicolumn{1}{c}{$ I^{\pi}_i  \rightarrow I^{\pi}_f$} &  \multicolumn{1}{c}{$a_2$} & \multicolumn{1}{c}{$a_4$} & \multicolumn{1}{c}{ $R_{ac}$} & \multicolumn{1}{c}{Mult.}\\ \hline 
\endhead
\hline\hline

\endfoot

\hline \hline
\endlastfoot

{Band D1} 	
&	134.3	&$	134.3	$&$	3/2^{+}	\rightarrow	1/2^{+}	$	&-0.32(2)	&0.04(1)	&0.87(3)	&	$M1/E2$	\\
&	183.6	&$	 317.9	$&$	5/2^{+}	\rightarrow	3/2^{+}	$	&-0.15(3)	&0.10(3)	&0.73(12)	&	$M1/E2$	\\
&	244.6	&$	1445.7	$&$	13/2^{+}	\rightarrow	11/2^{+}	$	&-0.22(2)	&0.07(3)	&0.77(4)*	&	$M1/E2$	\\
&	244.9	&$	815.4	$&$	9/2^{+}	\rightarrow	7/2^{+}	$	&-0.22(2)	&0.07(3)	&0.77(4)*	&	                     	\\
&	252.6	&$	570.5	$&$	7/2^{+}	\rightarrow	5/2^{+}	$	&-0.41(3)	&0.12(5)	&0.81(3)	&	$M1/E2$	\\
&	317.9	&$	317.9	$&$	5/2^{+}	\rightarrow	1/2^{+}	$	&0.42(2)	&0.22(3)	&1.32(2)	&	$E2$	\\
&	385.7	&$	1201.1	$&$	11/2^{+}	\rightarrow	9/2^{+}	$	&	&	&	&	            	 \\
&	436.2	&$	570.5	$&$	7/2^{+}	\rightarrow	3/2^{+}	$	&0.35(2)	&0.12(2)	&	&	$E2$	\\
&	497.5	&$	815.4	$&$	9/2^{+}	\rightarrow	5/2^{+}	$	&	&	&	&	       		  \\
&	630.3	&$	1445.7	$&$	13/2^{+}	\rightarrow	9/2^{+}	$	&	&	&	&	$E2$	\\
&	630.6	&$	1201.1	$&$	11/2^{+}	\rightarrow	7/2^{+}	$	&	&	&	&	     		\\
&	711.5	&$	2170.9	$&$	17/2^{+}	\rightarrow	13/2^{+}	$	&	&	&	&			\\
&	725.4	&$	2170.9	$&$	17/2^{+}	\rightarrow	13/2^{+}	$	&0.32(6)	&-0.16(7)	&1.31(11)	&	$E2$	\\
&	731.3	&$	1932.3	$&$	15/2^{+}	\rightarrow	11/2^{+}	$	&0.77(6)	&-0.26(5)	&1.25(2)	&	$E2$	\\
&	767.7	&$	1445.7	$&$	13/2^{+}	\rightarrow	  9/2^{+}	$	&0.44(9)	&0.03(12)	&1.30(11)	&	$E2$	\\
\hline																		
Band D2	
&	170.2	&$	207.4	$&$	11/2^{-}	\rightarrow	9/2^{-}	$	&-0.24(2)	&0.04(2)	&0.82(2)	&	$M1/E2$	\\
&	234.9	&$	827.1	$&$	15/2^{-}	\rightarrow	13/2^{-}	$	&-0.44(1)	&0.13(1)	&0.90(4)	&	$M1/E2$	\\
&	246.2	&$	1590.0	$&$	19/2^{-}	\rightarrow	17/2^{-}	$	&-0.32(3)	&0.06(3)	&0.88(3)	&	$M1/E2$	\\
&	286.2	&$	2486.1	$&$	23/2^{-}	\rightarrow	21/2^{-}	$	&	&	&	&	 	\\
&	302.9	&$	3432.6	$&$	27/2^{-}	\rightarrow	(25/2^{-})	$	&	&	&	&		 \\
&	384.8	&$	592.2	$&$	13/2^{-}	\rightarrow	11/2^{-}	$	&-0.35(2)	&0.08(3)	&0.82(5)	&	$M1/E2$	\\
&	516.7	&$	1343.8	$&$	17/2^{-}	\rightarrow	15/2^{-}	$	&-0.41(4)	&0.11(2)	&0.76(3)	&	$M1/E2$	\\
&	555.0	&$	592.2	$&$	13/2^{-}	\rightarrow	9/2^{-}	$	&0.54(2)	&-0.21(4)	&1.33(4)	&	$E2$	\\
&	609.4	&$	2199.9	$&$	21/2^{-}	\rightarrow	19/2^{-}	$	&	&	&	&		\\
&	619.7	&$	827.1	$&$	15/2^{-}	\rightarrow	11/2^{-}	$	&0.45(2)	&-0.11(3)	&1.23(2)	&	$E2$	\\
&	643.6	&$	3129.7	$&$	(25/2^{-})	\rightarrow	23/2^{-}	$	&	&	&	&		\\	
&	751.6	&$	1343.8	$&$	17/2^{-}	\rightarrow	13/2^{-}	$	&0.39(2)	&-0.23(2)	&1.34(2)	&	$E2$	\\
&	762.9	&$	1590.0	$&$	19/2^{-}	\rightarrow	15/2^{-}	$	&0.44(2)	&-0.14(2)	&1.22(3)	&	$E2$	\\
&	856.1	&$	2199.9	$&$	21/2^{-}	\rightarrow	17/2^{-}	$	&0.38(2)	&0.01(2)	&1.34(2)	&	$E2$	\\
&	896.1	&$	2486.1	$&$	23/2^{-}	\rightarrow	19/2^{-}	$	&0.41(4)	&-0.31(1)	&1.21(4)	&	$E2$	\\
&	929.8	&$	3129.7	$&$	(25/2^{-})	\rightarrow	21/2^{-}	$	&	&	&	&		 \\
&	946.5	&$	3432.6	$&$	27/2^{-}	\rightarrow	23/2^{-}	$	&0.28(3)	&-0.13(4)	&1.42(5)	&	$E2$	\\
&	970.6	&$	4403.2	$&$	(31/2^{-})	\rightarrow	27/2^{-}	$	&	&	&	&	 	\\
&	994.8	&$	5398.0	$&$	(35/2^{-})	\rightarrow	(31/2^{-})	$	&0.34(3)	&0.02(4)	&1.29(9)	&	$E2$	\\
&	1001.3	&$	4131.0	$&$	(29/2^{-})	\rightarrow	(25/2^{-})	$	&	&	&	&		\\
&	1110.1	&$	5513.6	$&$	(35/2^{-})	\rightarrow	(31/2^{-})	$	&0.46(6)	&0.17(8)	&1.28(10)	&	$E2$	\\	
\hline																		
{Band D3} 	
&	126.6	& $	2297.5	$&$	19/2^{+}	\rightarrow	17/2^{+}	$	&	&	&	&		\\
&	159.7	&$	2457.2	$&$	21/2^{+}	\rightarrow	19/2^{+}	$	&-0.26(2)	&0.06(3)	&0.79(4)	&	$M1/E2$	\\
&	164.3	&$	2096.6	$&$	17/2^{+}	\rightarrow	15/2^{+}	$	&	&	&	&		\\
&	189.7	&$	2646.9	$&$	23/2^{+}	\rightarrow	21/2^{+}	$	&-0.32(2)	&0.11(2)	&0.85(3)	&	$M1/E2$	\\
&	199.0	&$	2096.6	$&$	17/2^{+}	\rightarrow	15/2^{+}	$	&-0.27(1)	&0.11(2)	&0.89(6)	&	$M1/E2$	\\
&	200.9	&$	2297.5	$&$	19/2^{+}	\rightarrow	17/2^{+}	$	&-0.27(1)	&0.11(2)	&0.89(6)	&	$M1/E2$	\\
&	234.9	&$	2881.8	$&$	25/2^{+}	\rightarrow	23/2^{+}	$	&	&	&	&	$M1/E2$	\\
&	294.7	&$	3176.5	$&$	27/2^{+}	\rightarrow	25/2^{+}	$	&	&	&	&	$M1/E2$	\\
&	349.4	&$	2646.9	$&$	23/2^{+}	\rightarrow	19/2^{+}	$	&	&	&	&		\\
&	357.4	&$	3533.9	$&$	29/2^{+}	\rightarrow	27/2^{+}	$	&	&	&	&	$M1/E2$	\\
&	360.6	&$	2457.2	$&$	21/2^{+}	\rightarrow	17/2^{+}	$	&	&	&	&		\\
&	365.2	&$	2297.5	$&$	19/2^{+}	\rightarrow	15/2^{+}	$	&	&	&	&		\\
&	383.5	&$	3917.4	$&$	31/2^{+}	\rightarrow	29/2^{+}	$	&-0.62(3)	&0.08(3)	&0.87(5)	&	$M1/E2$	\\
&	399.9	&$	2297.5	$&$	19/2^{+}	\rightarrow	15/2^{+}	$	&	&	&	&	$E2$	\\
&	424.6	&$	2881.8	$&$	25/2^{+}	\rightarrow	21/2^{+}	$	&	&	&	&	$E2$	\\
&	451.9	&$	1897.6	$&$	15/2^{+}	\rightarrow	13/2^{+}	$	&	&	&	&		\\
&	455.8	&$	4830.7	$&$	35/2^{+}	\rightarrow	33/2^{+}	$	&-0.33(4)	&0.08(2)	&0.78(2)*	&	$M1/E2$	\\
&	457.5	&$	4374.9	$&$	33/2^{+}	\rightarrow	31/2^{+}	$	&-0.33(4)	&0.08(2)	&0.78(2)*	&	$M1/E2$  \\
&	506.6	&$	2096.6	$&$	17/2^{+}	\rightarrow	19/2^{-}	$	&	&	&0.78(3)	&$E1$		\\
&	510.4	&$	5876.6	$&$	39/2^{+}	\rightarrow	37/2^{+}	$	&	&	&0.98(4)	&$M1/E2$		\\
&	529.6	&$	3176.5	$&$	27/2^{+}	\rightarrow	23/2^{+}	$	&	&	&1.34(2)	&$E2$		\\
&	535.5	&$	5366.2	$&$	(37/2^{+})	\rightarrow	35/2^{+}	$	&	&	&0.82(4)	&$M1/E2$		\\
&	558.0	&$	7024.0	$&$	(43/2^{+})	\rightarrow	(41/2^{+})	$	&	&	&0.88(5)	&$M1/E2$		\\
&	595.2	&$	7619.2	$&$	(45/2^{+})	\rightarrow	(43/2^{+})	$	&	&	&0.78(7)	&$M1/E2$		\\
&	(620.6)	&$	(8239.8)	$&$	(47/2^{+})	\rightarrow	(45/2^{+})	$	&	&	&	&$M1/E2$		\\
&	637.3	&$	2096.6	$&$	17/2^{+}	\rightarrow	13/2^{-}	$	&	&	&	&		\\
&	650.9	&$	2096.6	$&$	17/2^{+}	\rightarrow	13/2^{+}	$	&	&	&	&	        \\
&	652.1	&$	3533.9	$&$	29/2^{+}	\rightarrow	25/2^{+}	$	&0.43(2)	&-0.21(3)	&1.43(4)	&$E2$		\\
&	696.5	&$	1897.6	$&$	15/2^{+}	\rightarrow	11/2^{+}	$	&	&	&	&		\\
&	740.9	&$	3917.4	$&$	31/2^{+}	\rightarrow	27/2^{+}	$	&	&	&	&	$E2$	\\
&	841.0	&$	4374.9	$&$	33/2^{+}	\rightarrow	29/2^{+}	$	&0.41(3)	&-0.21(4)	&1.32(1)&$E2$	\\
&	896.0	&$	2096.6	$&$	17/2^{+}	\rightarrow	15/2^{-}	$	&	&	&	&		\\
&	913.8	&$	4830.7	$&$	35/2^{+}	\rightarrow	31/2^{+}	$	&	&	&	&	$E2$	\\
&	953.7	&$	2297.5	$&$	19/2^{+}	\rightarrow	17/2^{-}	$	&-0.14(1)	&0.05(2)	&0.89(2)	&	$E1$	\\
&	991.3	&$	5366.2	$&$	(37/2^{+})	\rightarrow	33/2^{+}	$	&	&	&1.32(1)	&$E2$	\\
&	1045.9	&$	5876.6	$&$	(39/2^{+})	\rightarrow	35/2^{+}	$	&	&	&1.34(2)	&$E2$	\\
&	1099.8	&$	6466.0	$&$	(41/2^{+})	\rightarrow	(37/2^{+})	$	&0.43(2)	&-0.21(3)	&1.14(2)	&$E2$	\\
&	1147.4	&$	7024.0	$&$	(43/2^{+})	\rightarrow	(39/2^{+})	$	&	&	&1.42(1)	&$E2$		\\
&	1153.2	&$	7619.2	$&$	(45/2^{+})	\rightarrow	(41/2^{+})	$	&	&	&1.33(4)	&$E2$            \\
&	1215.8	&$	8239.8	$&$	(47/2^{+})	\rightarrow	(43/2^{+})	$	&	&	&	&$E2$	\\
&	1269.5	&$	2096.6	$&$	17/2^{+}	\rightarrow	15/2^{-}	$	&-0.12(3)	&0.05(6)	&0.78(4)	&	$E1$	\\
&	1305.4	&$	1897.6	$&$	15/2^{+}	\rightarrow	13/2^{-}	$	&-0.22(4)	&0.10(2)	&0.89(2)	&	$E1$	\\										\hline																		
Band D4	
&	118.0	&$	2621.6	$&$	21/2^{+}	\rightarrow	19/2^{+}	$	&	&	&	&	$M1/E2$	\\
&	129.7	&$	2416.1	$&$	19/2^{+}	\rightarrow	17/2^{+}	$	&	&	&0.95(6)	&	$M1/E2$	\\
&	205.5	&$	2621.6	$&$	21/2^{+}	\rightarrow	19/2^{+}	$	&-0.26(2)	&0.14(3)	&0.75(3)	&	$M1/E2$	\\
&	223.8	&$	2845.4	$&$	23/2^{+}	\rightarrow	21/2^{+}	$	&-0.43(2)	&0.13(3)	&0.64(8)	&	$M1/E2$	\\
&	283.6	&$	3129.0	$&$	25/2^{+}	\rightarrow	23/2^{+}	$	&	&	&0.76(5)	&	$M1/E2$	\\
&	305.6	&$	3434.6	$&$	27/2^{+}	\rightarrow	25/2^{+}	$	&	&	&0.57(8)	&	$M1/E2$	\\
&	319.5	&$	2416.1	$&$	19/2^{+}	\rightarrow	17/2^{+}	$	&	&	&	&		\\
&	324.1	&$	2621.6	$&$	21/2^{+}	\rightarrow	19/2^{+}	$	&	&	&	&		\\
&	346.4	&$	3781.0	$&$	29/2^{+}	\rightarrow	27/2^{+}	$	&	&	&0.76(3)	&	$M1/E2$	\\
&	388.2	&$	2845.4	$&$	23/2^{+}	\rightarrow	21/2^{+}	$	&	&	&	&		\\
&	429.3	&$	2845.4	$&$	23/2^{+}	\rightarrow	19/2^{+}	$	&	&	&1.32(3)	&$E2$		\\
&	431.7	&$	4212.7	$&$	31/2^{+}	\rightarrow	29/2^{+}	$	&	&	&0.76(3)	&	$M1/E2$	\\
&	446.0	&$	4658	$&$	33/2^{+}	\rightarrow	31/2^{+}	$	&	&	&0.93(5)	&	$M1/E2$	\\
&	482.1	&$	3129.0	$&$	25/2^{+}	\rightarrow	23/2^{+}	$	&	&	&0.76(3)	&$M1/E2$		\\
&	507.4	&$	3129.0	$&$	25/2^{+}	\rightarrow	21/2^{+}	$	&	&	&1.14(2)	&$E2$		\\
&	518.5	&$	2416.1	$&$	19/2^{+}	\rightarrow	15/2^{+}	$	&	&	&1.32(2)	&$E2$		\\
&	525.0	&$	2621.6	$&$	21/2^{+}	\rightarrow	17/2^{+}	$	&	&	&1.22(4)	&$E2$		\\
&	547.9	&$	2845.4	$&$	23/2^{+}	\rightarrow	19/2^{+}	$	&	&	&	&		\\
&	552.8	&$	3434.6	$&$	27/2^{+}	\rightarrow	25/2^{+}	$	&	&	&	&		\\
&	589.2	&$	3434.6	$&$	27/2^{+}	\rightarrow	23/2^{+}	$	&	&	&	&		\\
&	604.5	&$	3781.0	$&$	29/2^{+}	\rightarrow	27/2^{+}	$	&	&	&	&		\\
&	652.0	&$	3781.0	$&$	29/2^{+}	\rightarrow	25/2^{+}	$	&0.34(2)	&-0.23(4)	&1.32(4)	& $E2$		\\
&	671.8	&$	3129.0	$&$	25/2^{+}	\rightarrow	21/2^{+}	$	&	&	&1.34(3)	& $E2$	\\
&	778.1	&$	4212.7	$&$	31/2^{+}	\rightarrow	27/2^{+}	$	&	&	&	&		\\
&	787.7	&$	3434.6	$&$	27/2^{+}	\rightarrow	23/2^{+}	$	&	&	&	&	\\
&	826.1	&$	2416.1	$&$	19/2^{+}	\rightarrow	19/2^{-}	$	&	&	&	&		\\
&	877.8	&$	4658	$&$	33/2^{+}	\rightarrow	29/2^{+}	$	&	&	&1.43(4)	& $E2$		\\
&	899.2	&$	3781.0	$&$	29/2^{+}	\rightarrow	25/2^{+}	$	&	&	&	&		\\
&	1031.6	&$	2621.6	$&$	21/2^{+}	\rightarrow	19/2^{-}	$	&	&	&	&	$E1$	\\
&	1072.3	&$	2416.1	$&$	19/2^{+}	\rightarrow	17/2^{-}	$	&	&	&	&	$E1$	\\
&	1459.3	&$	2286.4	$&$	17/2^{+}	\rightarrow	15/2^{-}	$	&	&	&	&		\\	
\hline																												Band D5	
	&	215.5	& $	2959.5	$&$	23/2^{+}	\rightarrow	21/2^{+}	$	&	&	&0.72(5)	&	$M1/E2$	\\
	&	240.8	&$	2744	.0	$&$	21/2^{+}	\rightarrow	19/2^{+}	$	&-0.39(1)	&0.13(1)	&0.64(6)	&	$M1/E2$	\\
	&	276.7	&$	3236.2	$&$	25/2^{+}	\rightarrow	23/2^{+}	$	&-0.34(9)	&0.06(4)	&0.72(5)	&	$M1/E2$	\\
	&	286.8	&$	2744.0	$&$	21/2^{+}	\rightarrow	21/2^{+}	$	&	&	&	&		\\
	&	312.6	&$	2959.5	$&$	23/2^{+}	\rightarrow	23/2^{+}	$	&	&	&	&	\\
	&	327.9	&$	2744.0	$&$	21/2^{+}	\rightarrow	19/2^{+}	$	&-0.81(3)	&0.03(6)	&0.81(6)	&	$M1/E2$	\\
	&	335.4	&$	3571.6	$&$	27/2^{+}	\rightarrow	25/2^{+}	$	&-0.59(2)	&0.12(5)	&0.70(5)	&	$M1/E2$	\\
	&	337.9	&$	2959.5	$&$	23/2^{+}	\rightarrow	21/2^{+}	$	&-0.72(5)	&0.17(7)	&0.75(7)	&	$M1/E2$	\\
	&	354.4	&$	3236.2	$&$	25/2^{+}	\rightarrow	25/2^{+}	$	&	&	&	&	\\
	&	390.8	&$	3236.2	$&$	25/2^{+}	\rightarrow	23/2^{+}	$	&-0.82(5)	&0.24(4)	&0.79(5)	&	$M1/E2$	\\
	&	398.5	&$	4371.1	$&$	29/2^{+}	\rightarrow	27/2^{+}	$	&	&	&0.74(5)	&	$M1/E2$	\\
	&	401.0	&$	3972.6	$&$	31/2^{+}	\rightarrow	29/2^{+}	$	&	&	&0.71(4)	&	$M1/E2$	\\
	&	442.1	&$	4813.2	$&$	33/2^{+}	\rightarrow	31/2^{+}	$	&	&	&0.65(2)	&	$M1/E2$	\\
	&	446.5	&$	2744.0	$&$	21/2^{+}	\rightarrow	19/2^{+}	$	&	&	&	&		\\
	&	492.2	&$	3236.2	$&$	25/2^{+}	\rightarrow	21/2^{+}	$	&	&	&	&		\\
	&	502.3	&$	2959.5	$&$	23/2^{+}	\rightarrow	21/2^{+}	$	&	&	&	&		\\
	&	543.4	&$	2959.5	$&$	23/2^{+}	\rightarrow	19/2^{+}	$	&0.28(3)	&0.16(3)	&1.26(8)	&	$E2$	\\
	&	589.3	&$	3236.2	$&$	25/2^{+}	\rightarrow	23/2^{+}	$	&	&	&	&		\\
	&	612.1	&$	3571.6	$&$	27/2^{+}	\rightarrow	23/2^{+}	$	&	&	&	&	\\
	&	614.6	&$	3236.2	$&$	25/2^{+}	\rightarrow	21/2^{+}	$	&	&	&	&	\\
	&	736.4	&$	3972.6	$&$	29/2^{+}	\rightarrow	25/2^{+}	$	&	&	&	&		\\
	&	799.5	&$	4371.1	$&$	31/2^{+}	\rightarrow	27/2^{+}	$	&	&	&1.26(9)	&	$E2$	\\
	&	840.6	&$	4813.2	$&$	33/2^{+}	\rightarrow	29/2^{+}	$	&	&	&	&		\\
	&	1154.0	&$	2744.0	$&$	21/2^{+}	\rightarrow	19/2^{-}	$	&	&	&	&		\\
	&	1159.4	&$	2503.2	$&$	19/2^{+}	\rightarrow	17/2^{-}	$	&	&	&	&	\\	
\hline																			
Band D6		
	&	140.2	&$	3376.1	$&$	25/2^{-}	\rightarrow	23/2^{-}	$	&-0.33(4)	&0.11(4)	&0.69(3)	&$M1/E2$		\\
         &	154.8	&$	3530.9	$&$	27/2^{-}	\rightarrow	25/2^{-}	$	&-0.26(2)	&0.05(3)	&0.78(3)	&	$M1/E2$	\\
	&	240.7	&$	3771.6	$&$	29/2^{-}	\rightarrow	27/2^{-}	$	&-0.23(2)	&0.06(3)	&0.67(3)	&	$M1/E2$	\\
	&	246.4	&$	3376.1	$&$	25/2^{-}	\rightarrow	25/2^{-}	$	&-0.36(1)	&0.13(2)	&0.88(1)	&	$M1/E2$	\\
	&	294.9	&$	4066.5	$&$	31/2^{-}	\rightarrow	29/2^{-}	$	&-0.41(5)	&0.20(3)	&0.69(7)	&	$M1/E2$	\\
	&	339.0	&$	3771.6	$&$	29/2^{-}	\rightarrow	27/2^{-}	$	&	&	&0.78(4)	&$M1/E2$		\\
	&	342.1	&$	4408.6	$&$	33/2^{-}	\rightarrow	31/2^{-}	$	&-0.54(2)	&0.10(2)	&0.73(2)	&	$M1/E2$	\\
	&	391.2	&$	4799.8	$&$	35/2^{-}	\rightarrow	33/2^{-}	$	&-0.63(2)	&0.09(3)	&0.65(1)	&	$M1/E2$	\\
	&	401.2	&$	3530.9	$&$	27/2^{-}	\rightarrow	25/2^{+}	$	&	&	&0.89(4)	&$E1$		\\
	&	415.5	&$	5215.3	$&$	37/2^{-}	\rightarrow	35/2^{-}	$	&	&	&0.67(1)	&	$M1/E2$	\\
	&	453.4	&$	5668.7	$&$	39/2^{-}	\rightarrow	37/2^{-}	$	&	&	&0.71(2)	&	$M1/E2$	\\
	&	479.3	&$	6148.0	$&$	41/2^{-}	\rightarrow	39/2^{-}	$	&-0.56(4)	&0.14(5)	&0.68(2)	&	$M1/E2$	\\
	&	511.7	&$	6659.7	$&$	43/2^{-}	\rightarrow	41/2^{-}	$	&	&	&0.75(4)	&	$M1/E2$	\\
	&	535.6	&$	4066.5	$&$	31/2^{-}	\rightarrow	27/2^{-}	$	&	&	&1.34(3)	&$E2$		\\
	&	542.5	&$	7202.2	$&$	45/2^{-}	\rightarrow	43/2^{-}	$	&	&	&0.93(8)	&	$M1/E2$	\\
	&	572.4	&$	7774.7	$&$	47/2^{-}	\rightarrow	45/2^{-}	$	&	&	&0.73(2)	&	$M1/E2$	\\
	&	595.1	&$	3771.6	$&$	29/2^{-}	\rightarrow	27/2^{+}	$	&	&	&0.89(4)	& $E1$	\\
	&	602.9	&$	8377.6	$&$	(49/2^{-})	\rightarrow	47/2^{-}	$	&	&	&0.78(3)	&$M1/E2$		\\
	&	637.0	&$	4408.6	$&$	33/2^{-}	\rightarrow	29/2^{-}	$	&0.41(3)	&-0.21(2)	&1.33(2)	& $E2$	\\
	&	649.1	&$	3530.9	$&$	27/2^{-}	\rightarrow	25/2^{+}	$	&	&	&0.93(4)	& $E1$		\\
	&	729.2	&$	3376.1	$&$	25/2^{-}	\rightarrow	23/2^{+}	$	&	&	&0.87(5)	& $E1$		\\
	&	733.3	&$	4799.8	$&$	35/2^{-}	\rightarrow	31/2^{-}	$	&0.32(4)	&-0.21(3)	&1.25(2)	&$E2$		\\
	&	749.8	&$	3235.9	$&$	23/2^{-}	\rightarrow	23/2^{-}	$	&-0.31(3)	&0.14(3)	&0.83(5)	&	$M1$	\\
	&	806.7	&$	5215.3	$&$	37/2^{-}	\rightarrow	33/2^{-}	$	&0.41(3)	&0.13(4)	&1.41(3)	&$E2$		\\
	&	868.9	&$	5668.7	$&$	39/2^{-}	\rightarrow	35/2^{-}	$	&0.32(1)	&0.05(2)	&1.31(4)	&$E2$	\\
	&	890.0	&$	3376.1	$&$	25/2^{-}	\rightarrow	23/2^{-}	$	&	&	&0.78(3)	& $M1/E2$		\\
	&	932.7	&$	6148.0	$&$	41/2^{-}	\rightarrow	37/2^{-}	$	&	&	&1.42(1)	&$E2$	\\
	&	991.0	&$	6659.7	$&$	43/2^{-}	\rightarrow	39/2^{-}	$	&	&	&1.23(4)	&$E2$		\\
	&	1036.0	&$	3235.9	$&$	23/2^{-}	\rightarrow	21/2^{-}	$	&	&	&0.98(3)	&$M1/E2$	\\
	&	1044.8	&$	3530.9	$&$	27/2^{-}	\rightarrow	23/2^{-}	$	&	&	&1.27(4)	&$E2$		\\
	&	1054.2	&$	7202.2	$&$	45/2^{-}	\rightarrow	41/2^{-}	$	&0.41(3)	&-0.24(5)	&1.43(2)	&$E2$		\\
	&	1076.2	&$	3376.1	$&$	25/2^{-}	\rightarrow	21/2^{-}	$	&	&	&1.15(3)	&	$E2$	\\
	&	1115.0	&$	7774.7	$&$	47/2^{-}	\rightarrow	43/2^{-}	$	&	&	&1.24(3)	&$E2$	\\
	&	1175.4	&$	8377.6	$&$	(49/2^{-})	\rightarrow	45/2^{-}	$	&	&	&	&$E2$		\\
	&	1176.2	&$	3376.1	$&$	25/2^{-}	\rightarrow	21/2^{-}	$	&	&	&1.34(3)	&	$E2$	\\	
\hline																		
Band D7	
	&	282.8	&$	4482.5	$&$	31/2^{-}	\rightarrow	29/2^{-}	$	&-0.53(2)	&0.13(3)	&0.77(1)	&	$M1/$E2$$	\\
	&	363.2	&$	4845.7	$&$	33/2^{-}	\rightarrow	31/2^{-}	$	&	&	&	&		\\
	&	367.4	&$	5213.1	$&$	35/2^{-}	\rightarrow	33/2^{-}	$	&-0.61(4)	&0.10(2)	& 0.79(3)	&	$M1/$E2$$	\\
	&	423.2	&$	6078.0	$&$	39/2^{-}	\rightarrow	37/2^{-}	$	&	&	&0.77(4)	&	$M1/$E2$$	\\
	&	441.7	&$	5654.8	$&$	37/2^{-}	\rightarrow	35/2^{-}	$	&-0.59(3)	&0.06(4)	&0.65(2)	&	$M1/$E2$$	\\
	&	467.4	&$	6545.4	$&$	41/2^{-}	\rightarrow	39/2^{-}	$	&-0.80(5)	&0.11(3)	& 0.70(1)	&	$M1/$E2$$	\\
	&	491.1	&$	7036.5	$&$	43/2^{-}	\rightarrow	41/2^{-}	$	&-0.81(7)	&0.23(4)	&0.57(3)	&	$M1/$E2$$	\\
	&	558.6	&$	7595.1	$&$	45/2^{-}	\rightarrow	43/2^{-}	$	&	&	&0.75(3)	&	$M1/$E2$$	\\
	&	646.0	&$	4845.7	$&$	33/2^{-}	\rightarrow	29/2^{-}	$	&	&	&	&	\\
	&	668.8	&$	4199.7	$&$	29/2^{-}	\rightarrow	27/2^{-}	$	&	&	&0.90(2)	&	$M1/$E2$$	\\
	&	710.9	&$	4482.5	$&$	31/2^{-}	\rightarrow	29/2^{-}	$	&-0.62(1)	&0.02(4)	&0.55(2)	&	$M1/$E2$$	\\
	&	730.6	&$	5213.1	$&$	35/2^{-}	\rightarrow	31/2^{-}	$	&	&	&	&		\\
	&	779.2	&$	4845.7	$&$	33/2^{-}	\rightarrow	31/2^{-}	$	&-0.73(3)	&0.12(3)	&0.70(1)	&	$M1/$E2$$	\\
	&	804.5	&$	5213.1	$&$	35/2^{-}	\rightarrow	33/2^{-}	$	&	&	&0.76(3)	&	$M1/$E2$$	\\
	&	809.1	&$	5654.8	$&$	37/2^{-}	\rightarrow	33/2^{-}	$	&	&	&	&		\\
	&	855.0	&$	5654.8	$&$	37/2^{-}	\rightarrow	35/2^{-}	$	&	&	&0.92(5)	&	$M1/$E2$$	\\
	&	864.9	&$	6078.0	$&$	39/2^{-}	\rightarrow	35/2^{-}	$	&	&	&	&		\\
	&	890.6	&$	6545.4	$&$	41/2^{-}	\rightarrow	37/2^{-}	$	&	&	&	&		\\
	&	951.6	&$	4482.5	$&$	31/2^{-}	\rightarrow	27/2^{-}	$	&	&	&1.34(7)	&	$E2$	\\
	&	958.5	&$	7036.5	$&$	43/2^{-}	\rightarrow	39/2^{-}	$	&	&	&	&		\\
	&	1049.7 	&$	7595.1	$&$	45/2^{-}	\rightarrow	41/2^{-}	$	&	&	&1.31(4)	&	$E2$	\\
	&	1074.1	&$	4845.7	$&$	33/2^{-}	\rightarrow	29/2^{-}	$	& 0.19(3)	&-0.03(3)	&1.32(4)	&	$E2$	\\
	&	1146.6	&$	5213.1	$&$	35/2^{-}	\rightarrow	31/2^{-}	$	&	&	&1.13(8)	&	$E2$	\\	
\hline																		
Band D8 
	&	401.3	&$	5423.0	$&$	(37/2^{+})	\rightarrow	(35/2^{+})	$	&-0.67(11)	&0.10(14)	&0.75(10)	&	$M1/E2$	\\
	&	437.0	&$	5860.0	$&$	(39/2^{+})	\rightarrow	(37/2^{+})	$	&-0.39(6)	&0.20(8)	&0.68(5)	&	$M1/E2$	\\
	&	474.9	&$	6335.9	$&$	(41/2^{+})	\rightarrow	(39/2^{+})	$	&-0.67(3)	&0.16(4)	&0.66(4)	&	$M1/E2$	\\
	&	507.5	&$	6843	.4	$&$	(43/2^{+})	\rightarrow	(41/2^{+})	$	&-0.17(4)	&0.08(5)	&0.64(4)	&	$M1/E2$	\\
	&	543.8	&$	7387.2	$&$	(45/2^{+})	\rightarrow	(43/2^{+})	$	&-0.56(15)&0.46(20)	&0.73(4)	&	$M1/E2$	\\
	&	576.4	&$	7963.6	$&$	(47/2^{+})	\rightarrow	(45/2^{+})	$	&	&	&	&	\\
	&	599.4	&$	8563.0	$&$	(49/2^{+})	\rightarrow	(47/2^{+})	$	&	&	&0.53(13)	&	$M1/E2$	\\
	&	838.3	&$	5860.0	$&$	(39/2^{+})	\rightarrow	(35/2^{+})	$	&	&	&	&	\\
	&	911.9	&$	6335.9	$&$	(41/2^{+})	\rightarrow	(37/2^{+})	$	&0.30(7)	&0.35(9)	&1.16(31)	&	$E2$	\\
	&	982.4	&$	6843.4	$&$	(43/2^{+})	\rightarrow	(39/2^{+})	$	&	&	&	&		\\
	&	1051.3	&$	7387.2	$&$	(45/2^{+})	\rightarrow	(41/2^{+})	$	&0.41(11)	&-0.15(14)	&1.58(50)	&	$E2$	\\
	&	1104.3	&$	5021.7	$&$	(35/2^{+})	\rightarrow	31/2^{+}	$	&	&	&	&		\\
	&	1120.2	&$	7963.6	$&$	(47/2^{+})	\rightarrow	(43/2^{+})	$	&	&	&	&	\\
	&	1175.8	&$	8563.0	$&$	(49/2^{+})	\rightarrow	(45/2^{+})	$	&0.45(6)	&0.12(8)	&1.42(84)	&	$E2$	\\
	&	1275.0	&$	9838.0	$&$	(53/2^{+})	\rightarrow	(49/2^{+})	$	&	&	&	&		\\																
\hline																			
Band D9	
	&	302.9	&$	6125.0	$&$	39/2^{+}	\rightarrow 37/2^{+} $		&-0.52(3)	&0.07(4)	&0.68(5)	&	$M1/E2$	\\
	&	466.3	&$	5822.1	$&$	37/2^{+}		\rightarrow	35/2^{-}	$	&-0.53(5)	&0.17(7)	&0.68(5)	&	$E1$	\\
	&	479.5	&$	6604.5	$&$	41/2^{+}	\rightarrow	39/2^{+}	$	&-0.60(6)	&0.05(8)	&0.64(4)	&	$M1/E2$	\\
	&	590.4	&$	7194.9	$&$	43/2^{+}	\rightarrow	41/2^{+}	$	&-0.39(7)	&0.01(9)	&0.75(10)	&	$M1/E2$	\\
	&	590.8	&$	7785.7	$&$	45/2^{+}	\rightarrow	43/2^{+}	$	&-0.37(4)	&-0.03(5)	&0.73(4)	&	$M1/E2$	\\
	&	606.7	&$	8392.4	$&$	47/2^{+}	\rightarrow	45/2^{+}	$	&0.84(15)	&0.64(21)	&0.41(3)	&	$M1/E2$	\\
	&	663.5	&$	7858	.4	$&$	41/2^{+}	\rightarrow	43/2^{+}	$	&-0.39(18)&0.66(26)	&0.67(5)	&	$M1/E2$	\\
	&	795.9	&$	9188.3	$&$	49/2^{+}	\rightarrow	47/2^{+}	$	&0.21(4)	&-0.05(6)	&1.21(8)	&	$E2$	\\
	&	847.0	&$	5822.1	$&$	37/2^{+}	\rightarrow	33/2^{+}	$	&-0.49(7)	&0.57(10)	&0.88(7)	&	$M1/E2$	\\

\hline
																													
Other states	
	&	282.6 		&$	4512.7	$&$	31/2^{+}		\rightarrow	29/2^{+}	$	&-0.32(3)	&0.18(4)	&0.68(5)&	$M1/E2$	\\
	&	371.5		&$	3943.1	$&$	29/2^{+}		\rightarrow	27/2^{+}	$	&-0.55(4)	&0.08(5)	&0.63(4)	&	$M1/E2$	\\
	&	382.2		&$	4325.3	$&$	31/2^{+}		\rightarrow	29/2^{+}	$	&-0.52(6)	&-0.23(8)	&0.73(4)	&	$M1/E2$	\\
	&	457.8		&$	4783.1	$&$	35/2^{+}		\rightarrow	31/2^{+}	$	&	&	&1.20(11)	&	$E2$	\\
	&	462.4		&$	4975.1	$&$	35/2^{+}		\rightarrow	31/2^{+}	$	&0.27(4)	&0.02(5)	&1.22(10)	&	$E2$	\\	
	&	693.2	        &$	5523.9	$&$	37/2^{+}		\rightarrow	35/2^{+}	$	& -0.25(5)&-0.02(7)	&0.81(6)	&$M1/E2$		\\
	&	753.7	        &$	4325.3	$&$	31/2^{+}		\rightarrow	27/2^{+}	$	&	&	&	&		\\
	&	783.2		&$	5758.3	$&$	35/2^{+}		\rightarrow	33/2^{+}	$	&-0.95(16)&-0.02(20)&0.48(4)	&	$M1/E2$	\\
	&	793.5	        &$	5355.8	$&$	35/2^{-}		\rightarrow	(31/2^{-})	$	&	&	&	&		\\	
	&	952.6	        &$	5355.8	$&$	35/2^{-}		\rightarrow	(31/2^{-})	$	&	&	&	&		\\
	&	1027.1	        &$	6551.0	$&$	41/2^{+}		\rightarrow	37/2^{+}	$	&1.08(12)	&0.22(15)	& 2.51(21)	&	$E2$	\\
	&	1101.1		&$	4230.1	$&$	29/2^{+}		\rightarrow	25/2^{+}	$	&0.32(12)	&0.06(17)	&1.14(8)	&	$E2$ \\
	&	1129.7	        &$	4562.3	$&$	(31/2^{-})		\rightarrow	27/2^{-}	$	&	&	&	&		\\	
			
\hline																										
Band Q1	
	&	302.9	& $	340.1	$&$	7/2^{+}	\rightarrow	9/2^{-}	$	&-0.21(4)	&0.07(5)	&0.84(5)	&	$E1$	\\
	&	337.9	&$	678.0	$&$	9/2^{+}	\rightarrow	7/2^{+}	$	&-0.49(3)	&0.13(4)	&0.66(4)	&	$M1/E2$	\\
	&	470.6	&$	678.0	$&$	9/2^{+}	\rightarrow	11/2^{-}	$	&	&	&	&	\\
	&	781.4	&$  1459.4	$&$	13/2^{+}	\rightarrow 	 9/2^{+}	$	&	&	&	&		\\
	&	843.5	&$  2302.9	$&$	(17/2^{+})	\rightarrow	13/2^{+}	$	&	&	&	&		\\									\hline																			
Band Q2	
	&	423.6	&$	4798.5 	$&$	35/2^{+}	\rightarrow	 33/2^{+}	$	&-0.59(3)	&0.21(4)	&0.58(5)	&	$M1/E2$	\\
	&	437.1	&$	4812.0	$&$	35/2^{+}	\rightarrow	33/2^{+}	$	&-0.54(6)	&-0.03(8)	&0.64(4)	&	$M1/E2$	\\
	&	702.5	&$	4812.0	$&$	35/2^{+}	\rightarrow	31/2^{+}	$	& 0.37(9)	&-0.05(11)	&1.50(13)	&	$E2$	\\
	&	811.7	&$	5623.7	$&$	39/2^{+}	\rightarrow	35/2^{+}	$	&0.30(2)	&-0.10(3)	&1.31(8)	&	$E2$	\\
	&	825.2	&$	5623.7	$&$	39/2^{+}	\rightarrow	35/2^{+}	$	&0.33(3)	&-0.19(4)	&1.36(10)	&	$E2$	\\
	&	881.1	&$	4798.5	$&$	35/2^{+}	\rightarrow	31/2^{+}	$	&0.46(6)	&-0.12(8)	&1.55(10)	&	$E2$	\\
	&	894.6	&$	4812.0	$&$	35/2^{+}	\rightarrow	31/2^{+}	$	&0.34(4)	&0.14(5)	&1.22(9)	&	$E2$	\\
	&	927.9	&$	6551.6	$&$	43/2^{+}	\rightarrow	39/2^{+}	$	&0.24(2)	&0.00(3)	&1.19(8)	&	$E2$	\\
	&	1016.1	&$	7567.7	$&$	47/2^{+}	\rightarrow	43/2^{+}	$	&0.31(3)	&-0.10(4)	&1.27(8)	&	$E2$	\\
	&	1106.9	&$	8674.6	$&$	51/2^{+}	\rightarrow	47/2^{+}	$	&0.40(1)	&-0.13(2)	&1.44(10)	&	$E2$	\\
	&	1186.9	&$	9861.5	$&$	55/2^{+}	\rightarrow	51/2^{+}	$	&0.54(7)	&-0.31(10)&1.72(18)	&	$E2$	\\
	&	1269.4	&$  11130.9	$&$	59/2^{+}	\rightarrow	55/2^{+}	$	&0.34(13)	&0.09(18)	&1.10(10)	&	$E2$	\\	
\hline																		
Band Q3	
	&	1132.8	&$	11023.8	$&$	59/2^{+}	\rightarrow	55/2^{+}	$	& 0.43(7)	&-0.27(9)	&1.47(13)	&	$E2$	\\
	&	1216.4	&$	  9891.0	$&$	55/2^{+}	\rightarrow	51/2^{+}	$	&0.70(21)	&-0.30(26)	&1.68(24)	&	$E2$	\\
	&	1251.7	&$	12275.5	$&$	(63/2^{+})	\rightarrow	59/2^{+}	$	&	&	&	&		\\	
\hline																		
Band Q4		
	&	417.6	&$	  3793.7$&$	27/2^{-}	\rightarrow	25/2^{-}	$	&-0.56(6)	&-0.05(8)	&0.69(5)	&	$M1/E2$	\\
	&	422.9	&$	  3793.7$&$	27/2^{-}	\rightarrow	23/2^{-}	$	&0.05(2)	&-0.03(2)	&1.20(6)	&	$E2$	\\
	&	465.0	&$	  1808.8$&$	19/2^{-}	\rightarrow	17/2^{-}	$	&-0.49(2)	&0.09(3)	&0.65(4)	&	$M1/E2$	\\
	&	489.2	&$	  4282.9$&$	31/2^{-}	\rightarrow	27/2^{-}	$	&0.19(2)	&-0.05(2)	&1.18(6)	&	$E2$	\\
	&	511.3	&$	  4282.9$&$	31/2^{-}	\rightarrow	29/2^{-}	$	&-0.35(7)	&0.27(10)	&0.72(5)	&	$M1/E2$	\\
	&	582.2	&$	  1409.3$&$	17/2^{-}	\rightarrow	15/2^{-}	$	&-0.83(7)	&0.31(9)	&0.45(3)	&	$M1/E2$	\\
	&	608.0	&$	  1200.2$&$	15/2^{-}	\rightarrow	13/2^{-}	$	&	& 	&	&		\\
	&	608.6	&$	  1808.8$&$	19/2^{-}	\rightarrow	15/2^{-}	$	&	&	&	&	\\
	&	643.5	&$	  4926.4$&$	35/2^{-}	\rightarrow	31/2^{-}	$	&0.26(1)	&-0.09(2)	&1.25(7)	&	$E2$	\\
	&	710.0	&$	  2300.0$&$	21/2^{-}	\rightarrow	19/2^{-}	$	&-0.51(4)	&0.26(5)	&0.59(6)    &	$M1/E2$	\\
	&      731.1	&$ 	  2540.2$&$     19/2^{-}     \rightarrow         19/2^{-}    $       & 	&	&	&	\\
	&	762.7	&$	  3370.8$&$	23/2^{-}	\rightarrow	19/2^{-}	$	&0.21(3)	&-0.05(4)	&1.21(7)	&	$E2$	\\
	&	767.9	&$	  5694.3$&$	39/2^{-}	\rightarrow	35/2^{-}	$	&0.31(3)	&-0.06(4)	&1.29(7)	&	$E2$	\\
	&	831.5	&$	  3370.8$&$	23/2^{-}	\rightarrow	19/2^{-}	$	&0.08(1)	&0.02(1)	&1.06(11)	&	$E2$	\\
	&	847.3	&$	  3370.8$&$	23/2^{-}	\rightarrow	19/2^{-}	$	&0.36(5)	&0.23(7)	&1.21(7)	&	$E2$	\\
	&	884.7	&$	  3370.8$&$	23/2^{-}	\rightarrow	23/2^{-}	$	&	&	&	&		\\
	&	888.4	&$	  6582.7$&$	43/2^{-}	\rightarrow	39/2^{-}	$	&0.33(3)	&-0.11(4)	&1.32(7)	&	$E2$	\\
	&	933.5	&$	  2523.5$&$	19/2^{-}	\rightarrow	19/2^{-}	$	&	&	&	&		\\
	&	956.2	&$	  2300.0$&$	21/2^{-}	\rightarrow	17/2^{-}	$	&	&	&1.16(54)	&	$(E2)$	\\
	&	981.7	&$	  1808.8$&$	19/2^{-}	\rightarrow	15/2^{-}	$	&0.18(4)	&-0.13(5)	&1.22(9)	&	$E2$	\\
	&      1000.1	&$	  7582.8$&$	47/2^{-}	\rightarrow	43/2^{-}	$	&0.35(4)	&-0.07(6)	&1.36(8)	&	$E2$	\\
	&	1018.1	&$	  2608.1$&$	19/2^{-}	\rightarrow	19/2^{-}	$	&0.35(3)	&0.03(12)	&1.26(13)	&	$M1/E2$ $(\Delta I=0)$	\\
	&	1070.8	&$	  3370.8$&$	23/2^{-}	\rightarrow	21/2^{-}	$	&-0.17(8)	&0.26(11)	&0.77(6)	&	$M1/E2$	\\
	&	1097.8	&$	  8680.6$&$	51/2^{-}	\rightarrow	47/2^{-}	$	&0.40(7)	&-0.13(9)	&1.43(12)	&	$E2$	\\
	&	1115.3	&$	  2523.5$&$	19/2^{-}	\rightarrow	17/2^{-}	$	&	&	&	&		\\
	&	1170.9	&$	  3370.8$&$	23/2^{-}	\rightarrow	21/2^{-}	$	&-0.56(11)	&0.20(15)	&0.64(6)	&	$M1/E2$	\\
	&	1179.7	&$	  2523.5$&$	19/2^{-}	\rightarrow	17/2^{-}	$	&-0.59(8)	&0.02(10)	&0.65(5)	&	$M1/E2$	\\
	&	1187.9	&$	  9868.5$&$	55/2^{-}	\rightarrow	51/2^{-}	$	&0.54(11)	&-0.07(15)	&1.50(16)	&	$E2$	\\
	&      1195.9	&$ 	  2540.2$&$     19/2^{-}     \rightarrow         17/2^{-}    $       & 	&	&	&	\\
	&	1264.3	&$  	  2608.1$&$	19/2^{-}	\rightarrow	17/2^{-}	$	&-0.11(8)	&0.13(7)	&0.81(6)	&	$M1/E2$	\\
	&	1274.6	&$	11143.1$&$	59/2^{-}	\rightarrow	55/2^{-}	$	&0.22(18)	&-0.19(24)&1.13(8)	&	$E2$	\\
	&	1356.1	&$	12499.2$&$	63/2^{-}	\rightarrow	59/2^{-}	$	&0.45(8)	&0.00(10)	&1.31(11)	&	$E2$	\\
	&	1424.9	&$	13924.1$&$	67/2^{-}	\rightarrow	63/2^{-}	$	&0.40(7)	&0.16(9)	&1.24(54)	&	$E2$	\\
	&	1500.8	&$	15424.9$&$	71/2^{-}	\rightarrow	67/2^{-}	$	&0.94(16)	&0.22(20)	&2.11(85)	&	$E2$	\\		
\hline																			
Band Q5	
	&	477.1	&$	  5628.8$&$	37/2^{(-)}	\rightarrow	33/2^{(-)}	$	&0.13(8)	&-0.33(10)&1.15(9)	&	$E2$	\\
	&	702.4	&$	  5628.8$&$	37/2^{(-)}	\rightarrow	35/2^{-}	$	&-0.16(5)	&0.08(7)	&0.86(6)	&	$M1/E2$	\\
	&	818.3	&$	  6512.6$&$	41/2^{(-)}	\rightarrow	39/2^{-}	$	&-0.28(4)	&0.02(6)	&0.78(5)	&	$M1/E2$	\\
	&	868.8	&$	  5151.7$&$	33/2^{(-)}	\rightarrow	31/2^{-}	$	&-0.30(11)	&0.24(16)	&0.65(5)	&	$M1/E2$	\\
	&	881.7	&$	  7394.3$&$	45/2^{(-)}	\rightarrow	41/2^{(-)}	$	&0.29(9)	&-0.16(12)&1.22(9)	&	$E2$	\\
	&	883.8	&$	  6512.6$&$	41/2^{(-)}	\rightarrow	37/2^{(-)}	$	&0.27(13)	&-0.01(8)	&1.20(8)	&	$E2$	\\
	&	890.5	&$	  8284.8$&$	49/2^{(-)}	\rightarrow	45/2^{(-)}	$	&0.32(4)	&-0.21(5)	&1.34(9)	&	$E2$	\\
	&	996.7	&$	  9281.5$&$	53/2^{(-)}	\rightarrow	49/2^{(-)}	$	&0.08(4)	&-0.12(6)	&1.06(8)	&	$E2$	\\
	&	1097.1	&$	10378.6$&$	57/2^{(-)}	\rightarrow	53/2^{(-)}	$	&0.41(7)	&-0.9(10)	&1.40(10)	&	$E2$	\\
	&	1134.4	&$	12878.3$&$	65/2^{(-)}	\rightarrow	61/2^{-}	$	&	&	&	&		\\
	&	1188.4	&$	11567.0$&$	61/2^{(-)}	\rightarrow	57/2^{(-)}	$	&	&	&1.08(21)	&	$E2$	\\
	&	1234.3	&$	  5151.7$&$	33/2^{(-)}	\rightarrow	31/2^{+}	$	&	&	&	&		\\
	&	1311.3	&$	12878.3$&$	65/2^{(-)}	\rightarrow	61/2^{(-)}	$	&	&	&	&		\\
	&	1378.0	&$	14256.3$&$	69/2^{(-)}	\rightarrow	65/2^{(-)}	$	&0.38(14)	&0.11(20)	&1.38(13)	&	$E2$	\\
	&	1485.5	&$	15741.8$&$	(73/2^{-})	\rightarrow	69/2^{(-)}	$	&	&	&	&		\\						
	&	1500.1	&$	15756.4$&$	(73/2^{-})	\rightarrow	69/2^{(-)}	$	&	&	&	&		\\
			\hline
																				
Band Q6
	&	978.8	&$	  9725.3	$&$	53/2^{-}	\rightarrow	49/2^{-}	$	&0.33(7)	&-0.01(9)	&1.29(10)	&	$E2$	\\
	&	990.0	&$	11753.9	$&$	61/2^{-}	\rightarrow	57/2^{-}	$	&0.65(9)	&0.23(2)	&1.49(12)	&	$E2$	\\
	&	1028.6	&$	10743.9	$&$	57/2^{-}	\rightarrow	53/2^{-}	$	&0.68(10)	&-0.15(13)&1.97(18)	&	$E2$	\\
	&	1057.8	&$	12800.7	$&$	65/2^{(-)}	\rightarrow	61/2^{(-)}	$	&0.35(14)	&0.17(19)	&1.32(12)	&	$E2$	\\
	&	1194.5 	&$	13996.2	$&$	69/2^{(-)}	\rightarrow	65/2^{(-)}	$	&0.40(16)	&-0.07(21)&1.49(17)	&	$E2$	\\
	&	1210.0	&$	12800.7	$&$	65/2^{(-)}	\rightarrow	(61/2^{-})	$	&	&	&	&	\\
	&	1213.1	&$	11591.7	$&$	(61/2^{-})	\rightarrow	57/2^{(-)}	$	&	&	&	&		\\	
	&	1234.7	&$	12800.7	$&$  65/2^{(-)}	\rightarrow	61/2^{(-)}	$	&0.45(10)	&0.05(14)	&1.35(17)	&	$E2$	\\
	&	1289.7	&$	15285.9	$&$  73/2^{(-)}	\rightarrow	69/2^{(-)}	$	&0.40(16)	&-0.07(21)&1.49(17)	&	$E2$	\\
	&	1393.8	&$	16679.7	$&$	(77/2^{-})	\rightarrow	73/2^{(-)}	$	&	&	&	&		\\
	&	1500.2	&$	18179.9	$&$	(81/2^{-})	\rightarrow	(77/2^{-})	$	&0.52(12)	&-0.24(16)&1.57(20)	&	$E2$	\\
	&	1603.9	&$	19783.8	$&$	(85/2^{-})	\rightarrow	(81/2^{-})	$	&	&	&	&		\\
\hline
																					
Band Q7	
	&	702.7	&$	  5848.4$&$	39/2^{-}	\rightarrow	35/2^{-}	$	&0.67(16)	&0.17(8)	&1.61(13)	&	$E2$	\\
	&	739.1	&$	  5147.7$&$	35/2^{-}	\rightarrow	33/2^{-}	$	&	&	&	&		\\
	&	771.8	&$	  6620.2$&$	43/2^{-}	\rightarrow	39/2^{-}	$	&0.42(4)	&-0.08(5)	&1.39(10)	&	$E2$	\\
	&	864.8	&$	  5147.7$&$	35/2^{-}	\rightarrow	31/2^{-}	$	&	&	&	&		\\
	&	880.9	&$	  7501.1$&$	47/2^{-}	\rightarrow	43/2^{-}	$	&0.35(3)	&0.00(4)&1.29(9)	&	$E2$	\\
	&	922.0	&$	  5848.4$&$	39/2^{-}	\rightarrow	35/2^{-}	$	&0.55(5)	&0.05(7)&	1.47(11)	&	$E2$	\\
	&	986.4	&$	  8487.5$&$	51/2^{-}	\rightarrow	47/2^{-}	$	&0.33(6)	&-0.20(8)	&1.39(9)	&	$E2$	\\
	&	1079.7	&$	  9567.2$&$	55/2^{-}	\rightarrow	51/2^{-}	$	&0.40(3)	&-0.09(4)	&1.38(9)*	&	$E2$	\\
	&	1081.2	&$	  5147.7$&$	35/2^{-}	\rightarrow	31/2^{-}	$	&0.40(3)	&-0.09(4)	&1.38(9)*	&	$E2$	\\
	&	1168.4	&$	10735.6$&$	59/2^{-}	\rightarrow	55/2^{-}	$	&0.44(7)	&-0.19(9)	&1.47(11)	&	$E2$	\\
	&	1242.4	&$	11978.0$&$	63/2^{-}	\rightarrow	59/2^{-}	$	&0.38(7)	&-0.20(9)	&1.51(13)	&	$E2$	\\
	&	1385.0	&$	13363.0$&$	(67/2^{-})	\rightarrow	63/2^{-}	$	&	&	&	&	\\
	&	1459.6	&$	14822.6$&$	(71/2^{-})	\rightarrow	(67/2^{-})	$	&	&	&	&	\\	
									
\hline

Band Q8		
	&	1172.1	&$	13150.1$&$	(65/2^{-})	\rightarrow	(63/2^{-})	$	&	&	&	&		\\
	&	1222.8	&$	14372.9$&$	(69/2^{-})	\rightarrow	(65/2^{-})	$	&	&	&	&		\\
	&	1300.7	&$	17010.3$&$	(77/2^{-})	\rightarrow	(73/2^{-})	$	&	&	&	&		\\
	&	1319.2	&$	18329.5$&$	(81/2^{-})	\rightarrow	(77/2^{-})	$	&	&	&	&		\\
	&	1336.7	&$	15709.6$&$	(73/2^{-})	\rightarrow	(69/2^{-})	$	&	&	&	&		\\
	&	1405.4	&$	19734.9$&$	(85/2^{-})	\rightarrow	(81/2^{-})	$	&	&	&	&		\\
	&	1507.2	&$	21242.1$&$	(89/2^{-})	\rightarrow	(85/2^{-})	$	&	&	&	&		\\
	&	1586.0	&$	22828.1$&$	(93/2^{-})	\rightarrow	(89/2^{-})	$	&	&	&	&		\\	
\hline
Band Q9		

	&	698.7	&$	  7284.3$&$	43/2^{(-)}	\rightarrow	(39/2^{-})	$	&	&	&	&		\\
	&	757.7	&$	  7284.3$&$	43/2^{(-)}	\rightarrow	41/2^{(-)}	$	&-0.76(6)	&0.15(8)	&0.49(4)	&	$M1/E2$	\\
	&	775.2	&$	  7284.3$&$	43/2^{(-)}	\rightarrow	39/2^{(-)}	$	&0.49(5)	&0.07(6)	&1.22(13)	&	$E2$	\\
	&	859.8	&$	  8144.1$&$	47/2^{(-)}	\rightarrow	43/2^{(-)}	$	&0.20(2)	&-0.16(2)	&1.23(9)	&	$E2$	\\
	&	897.8	&$	  6526.6$&$	41/2^{(-)}	\rightarrow	37/2^{(-)}	$	&0.38(3)	&-0.04(4)	&1.37(9)	&	$E2$	\\
	&	965.2	&$   	  9109.3$&$	51/2^{(-)}	\rightarrow	47/2^{(-)}	$	&0.45(6)	&-0.14(8)	&1.46(10)	&	$E2$	\\
	&	1001.2	&$	  6585.6$&$	(39/2^{-})	\rightarrow	35/2^{(-)}	$	&	&	&	&		\\
	&	1060.8	&$	10169.9$&$	55/2^{(-)}	\rightarrow	51/2^{(-)}	$	&0.30(6)	&-0.13(8)	&1.33(9)	&	$E2$	\\
	&	1150.4	&$	11320.4$&$	59/2^{(-)}	\rightarrow	55/2^{(-)}	$	&0.44(8)	&0.05(10)	&1.31(14)	&	$E2$	\\
	&	1244.2	&$	12564.3$&$	63/2^{(-)}	\rightarrow	59/2^{(-)}	$	&0.46(10)	&-0.29(13)&1.59(16)	&	$E2$	\\
	&	1339.6	&$	13904.3$&$	67/2^{(-)}	\rightarrow	63/2^{(-)}	$	&0.22(2)	&0.04(3)	&1.19(9)	&	$E2$	\\
	&	1436.1	&$	15340.4$&$	71/2^{(-)}	\rightarrow	67/2^{(-)}	$	&0.41(14	&0.13(19)	&1.26(13)	&	$E2$	\\
	&	1452.6	&$	16793.0$&$	(75/2^{-})	\rightarrow	71/2^{(-)}	$	&	&	&	&		\\
	&	1497.4	&$	18290.2$&$	(79/2^{-})	\rightarrow	(75/2^{-})	$	&	&	&	&		\\
									
\hline
Band Q10 	
	&	1158.1	&$	11328.0	$&$	59/2^{(-)}	\rightarrow	55/2^{(-)}	$	&0.45(5)	&0.05(6)	&1.35(12)	&	$E2$	\\
	&	1199.3	&$	12592.3	$&$	63/2^{(-)}	\rightarrow	59/2^{(-)}	$	&0.43(6)	&-0.28(7)	&1.59(17)	&	$E2$	\\
	&	1240.5	&$	13767.8	$&$	67/2^{(-)}	\rightarrow	63/2^{(-)}	$	&0.69(14)	&0.16(18)	&1.50(20)	&	$E2$	\\
	&	1274.7	&$	15042.5	$&$	71/2^{(-)}	\rightarrow	67/2^{(-)}	$	&0.30(6)	&0.18(8)	&1.19(12)	&	$E2$	\\
	&	1374.9	&$	16417.4	$&$	75/2^{(-)}	\rightarrow	71/2^{(-)}	$	&0.57(7)	&0.03(9)	&1.45(16)	&	$E2$	\\
	&	1484.8	&$	17902.2	$&$	(79/2^{-})	\rightarrow	75/2^{(-)}	$	&	&	&	&		\\
	&	1592.7	&$	19494.9	$&$	(83/2^{-})	\rightarrow	(79/2^{-})	$	&	&	&	&		\\
	&	1699.1	&$	21194.0	$&$	(87/2^{-})	\rightarrow	(83/2^{-})	$	&	&	&	&		\\								\hline																																							
Band Q11	
	&	700.6	&$	4740.7$&$	31/2^{(-)}	\rightarrow	27/2^{(-)}	$	&	&	&1.31(10)	&	$E2$	\\
	&	843.7	&$	5584.5$&$	35/2^{(-)}	\rightarrow	31/2^{(-)}	$	&	&	&1.12(10)	&	$E2$	\\
	&	924.7	&$	6509.1$&$	39/2^{(-)}	\rightarrow	35/2^{(-)}	$	&0.13(6)	&-0.19(9)	&1.18(9)	&	$E2$	\\																
 \end{longtable*}
\end{center}
\renewcommand{\thefootnote}{\arabic{footnote}}
 
\begin{center}
\begin{figure*}[t!]
\includegraphics[width=0.8\textwidth]{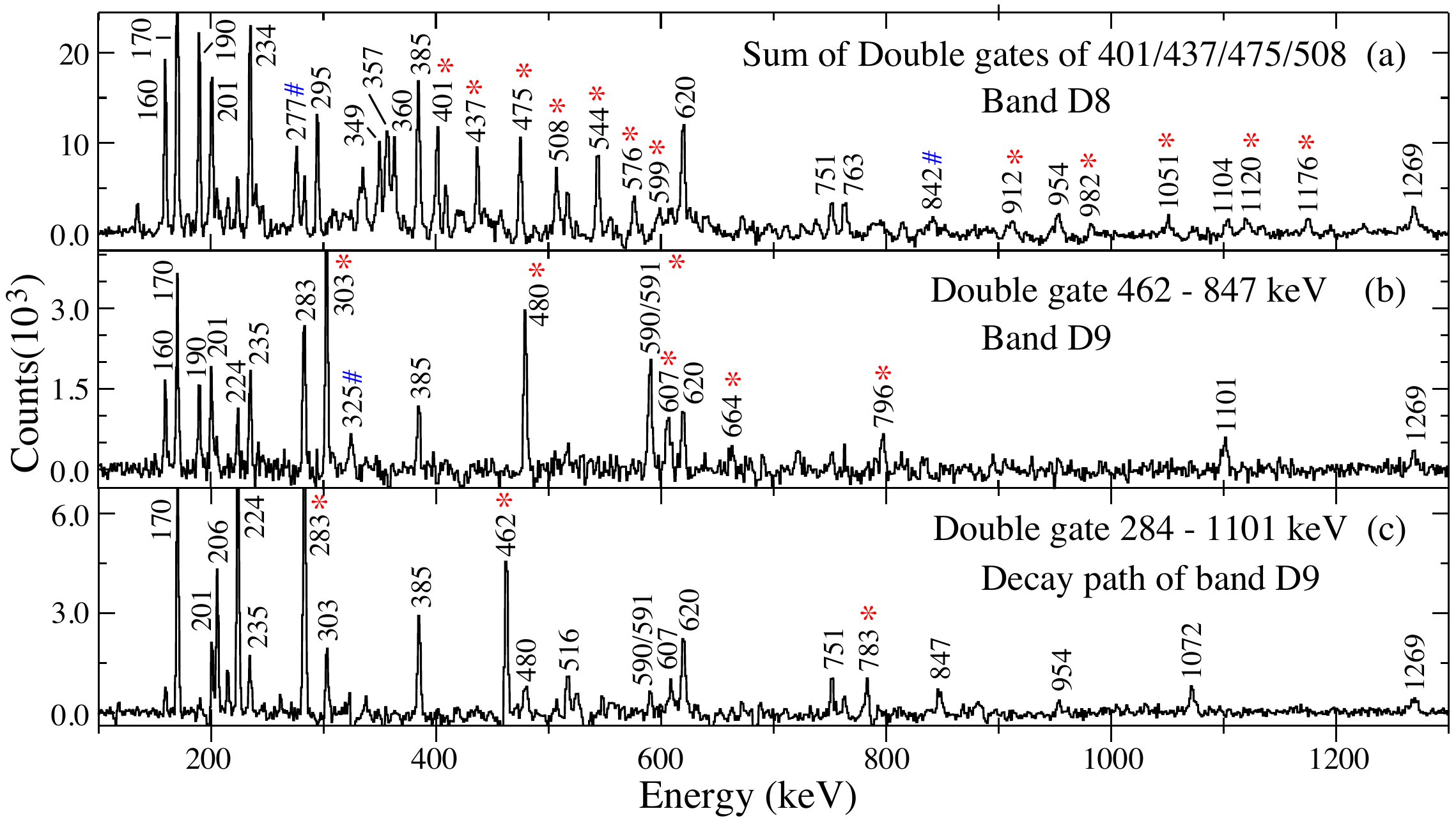}
\caption{\label{fig3spe} (Color online) Representative double-gated coincidence spectra for bands $D8$ and $D9$ in $^{133}$Ce. The gates were placed on selected transitions in each sequence and are indicated. The transitions marked with asterisks represent the members of the band, while those marked with \# are identified contaminants from other bands.}
\end{figure*}
\end{center}

\subsection{\label{sec-dipole} The dipole bands } %
In addition to the dipole bands ($D3$ - $D7$) previously identified in Refs.~\cite{133ce-a,ma1987}, two new sequences of dipole character, labeled as $D8$ and $D9$ in Fig.~\ref{FIG1}, have been observed. Spectra obtained by double-gating on selected transitions of each of these bands are displayed in Fig.~\ref{fig3spe}. 

Band $D8$ [Fig.~\ref{fig3spe}(a)]  is composed of nine levels linked by dipole and quadrupole crossover transitions. It decays through the 1104-keV transition and feeds into band $D3$ at the 31/2$^+$ level. While the bandhead energy of band $D8$ is fixed at 5021.7 keV, its spin-parity $(35/2^+)$ is tentative, since it was not possible to obtain either an angular distribution or a correlation ratio for this 1104-keV transition. The present tentative assignment is based on comparisons with similar dipole bands and the theoretical interpretation of the band (see Section~\ref{sec-dis}). 

The dipole sequence, $D9$ [Fig.~\ref{fig3spe}(b)] is comprised of six states connected by mixed dipole-quadrupole transitions. It decays primarily through the 847-keV, $E2$ transition towards the 4975.1-keV state, which in turn decays to band $D4$ through the 462-283-1101 keV cascade [Fig.~\ref{fig3spe}(c)]. Band $D9$ also decays very weakly through the 466-keV $\gamma$ ray towards the 5355.8-keV state. The latter in turn is deexcited toward band $D2$ via the 794-, 953- and 1130-keV transitions. Two other lines of 664 and 796 keV were identified in coincidence with transitions of band $D9$, which depopulate the $45/2^+$ and $51/2^+$ states at 7858.4 and 9188.3 keV, respectively. The bandhead of $D9$, populated by the 303-keV transition, is fixed at 5822.1 keV and has spin-parity $37/2^+$. 

Three new states linked by the 382- and 458-keV transitions and connected to the 27/2$^+$ level of band $D5$ by the 372- and 754-keV $\gamma$ rays were also identified. The  382- and 458-keV transitions have different character, being dipole and quadrupole transitions, respectively. Therefore, this sequence is not considered to be a band. 
 
\subsection{The quadrupole bands}
\begin{center}
\begin{figure*}[t!]
\includegraphics[width=0.8\textwidth]{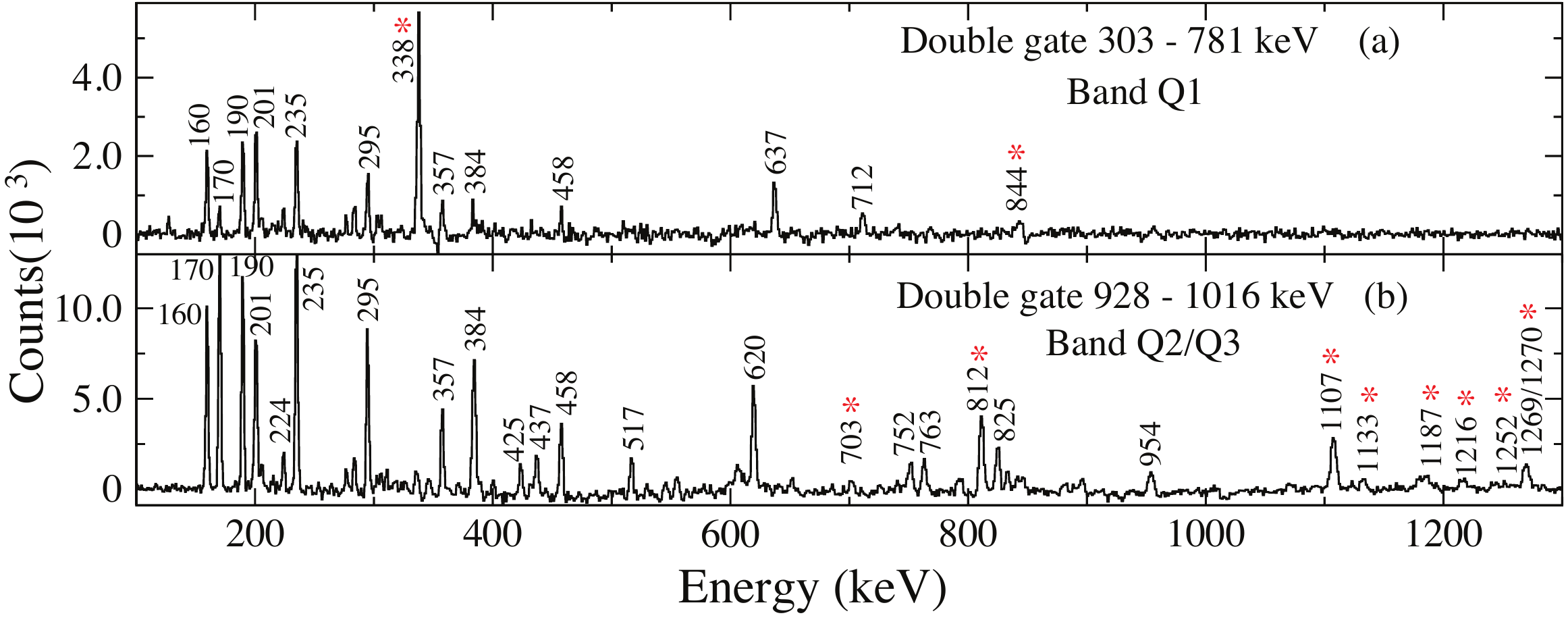}
\caption []{\label{Q1} (Color online) Representative double-gated coincidence spectra for  bands $Q1$, $Q2$ and $Q3$ in $^{133}$Ce. The gates were placed on selected transitions in each band. Transitions marked with asterisks represent the band members, while those marked with \# are identified contaminants from other bands.} 
\end{figure*}
 \end{center}

\begin{figure*}[]
\includegraphics[width=0.79\textwidth]{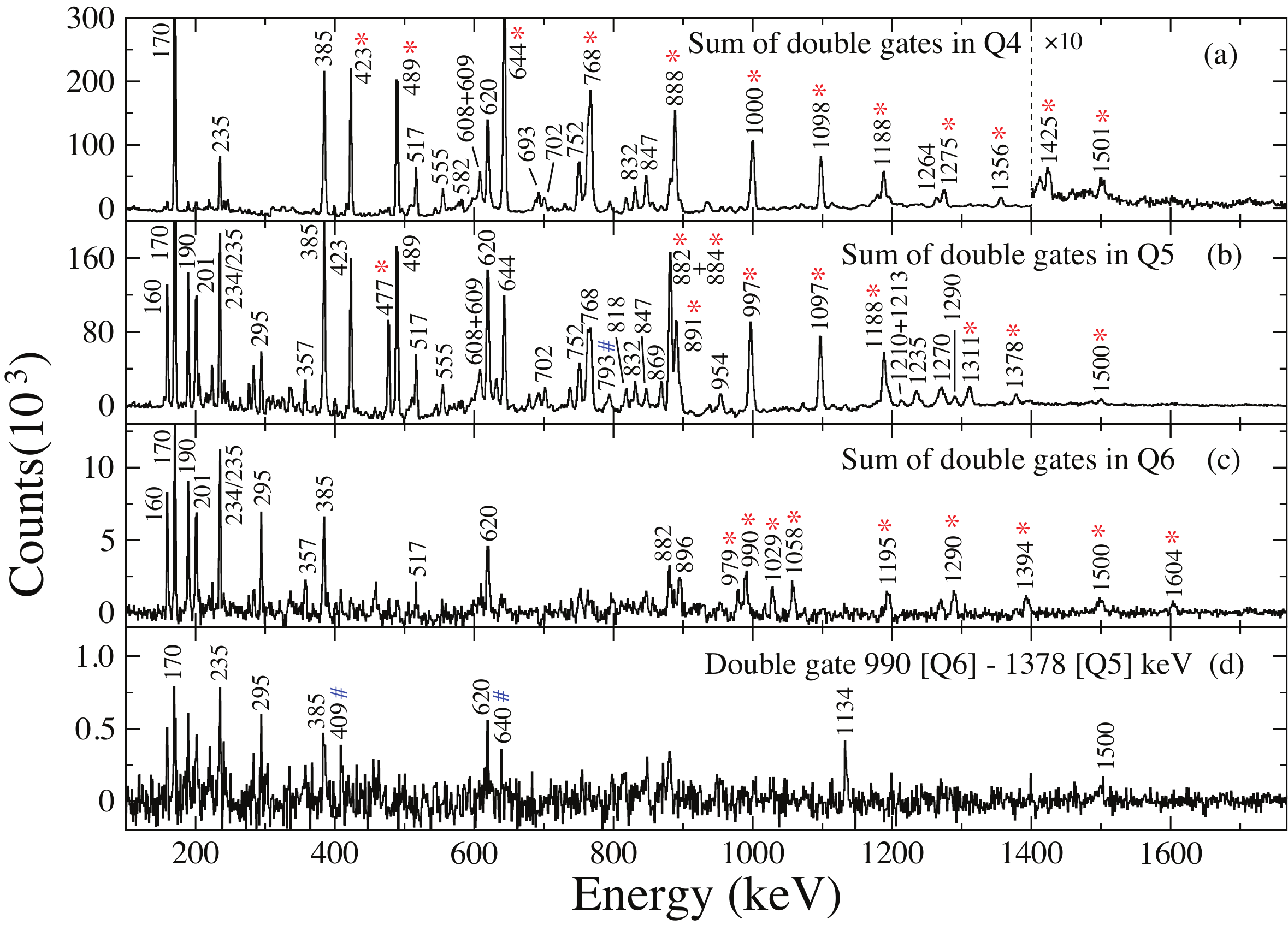}
\caption []{\label{Q2} (Color online) Representative double-gated coincidence spectra for  bands $Q4$, $Q5$ and $Q6$ in $^{133}$Ce. The gates were set on selected transitions in each band. The transitions marked with asterisks represent the band members, while those marked with \# are identified contaminants from other bands.} 
\end{figure*}

\begin{figure*}[]
\includegraphics[width=0.79\textwidth]{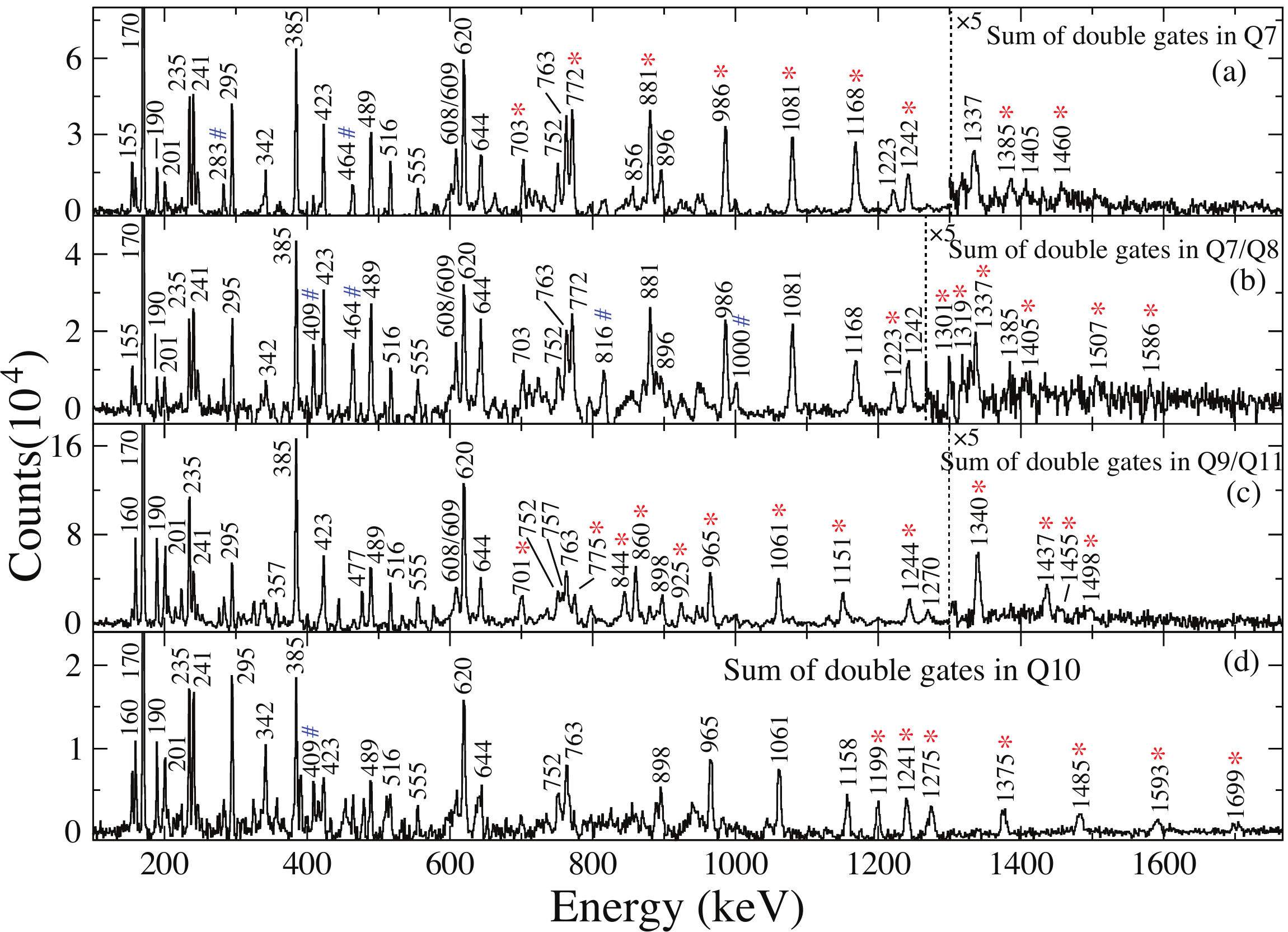}
\caption []{\label{Q3} (Color online) Representative double-gated coincidence spectra for  bands $Q7$, $Q8$, $Q9$, $Q10$ and $Q11$ in $^{133}$Ce. The gates were set on selected transitions in each band. The transitions marked with asterisks represent the band members, while those marked with \# are identified contaminants from other bands.} 
\end{figure*}

Two $\Delta I =2$ bands, $Q1$ (Fig.~\ref{FIG1}) and $Q11$ (Fig.~\ref{FIG2}), with positive and negative parity, respectively, have been observed at low and medium spins, along with several new states which are fed by transitions from band $Q4$ (Fig.~\ref{FIG2}). In addition, two new sequences of quadrupole character, $Q2$ and $Q3$ (Fig.~\ref{FIG1}), have been identified at high spins. The relative spins of these bands were established from measurements of both angular distributions and correlations as described in Sec.~\ref{sec-exp}. Spectra obtained by double-gating on selected $\gamma$ rays in each of these bands are presented in Figs. \ref{Q1}, \ref{Q2}, and \ref{Q3}. 

The quadrupole sequence $Q1$ is composed of four states observed for the first time in this study. The 303- and 338-keV transitions are of dipole character, fixing the spins of the lowest two levels as 7/2 and 9/2, respectively. The positive parity assigned to the $9/2$, 678.0-keV state comes from the presence of the 768-keV feeding transition of $E2$ character linking it to band $D1$. The $7/2^+$ assignment to the bandhead then follows from the dipole-quadrupole mixed character of the 338-keV $\gamma$ ray, while the $13/2^+$ spin-parity of the third member in the sequence is based on the likely $E2$ character of the 712-keV link with the $17/2^+$ state in band $D1$. 

The decay sequence labeled $Q2$ in Fig.~\ref{FIG1} is entirely new. It consists of eight levels connected by transitions of $E2$ character and decays through five transitions towards band $D3$. While the decay of the $31/2^+$ bandhead at 4109.5 keV could not be delineated, the excitation energy, spins and positive parity of the band are firmly established by the connecting transitions to the dipole sequence, $D3$. Moreover, band $Q3$, which is also new, decays to band $Q2$ through the 1216-keV $E2$ transition. Therefore, the bandhead of band $Q3$ at 9891.0 keV has spin-parity $55/2^+$.  

In addition, all transitions assigned to bands designated as $2-4$ in Ref.~\cite{karl1996} have been observed in the present work and their placement is consistent with the proposed level scheme, albeit, with some modifications. Band $Q4$ was first reported in Ref. \cite{ma1987} (as band 1) and was later extended to higher spins in Ref. \cite{karl1996}, but was, however, never linked to low-lying states; four transitions (832, 847, 934 and 1180 keV) were observed in coincidence with band $Q4$ and suggested as possible links~\cite{karl1996} to the low-lying structure. In the present work, these transitions, along with several others, have been used to link band $Q4$ to the low-lying states (see Fig.~\ref{FIG2}). This resulted in a bandhead spin-parity  of  $23/2^-$, which is $2 \hbar$ lower than the previously assumed value of $27/2$ in Ref.~\cite{karl1996}. All other previously observed in-band transitions are confirmed. The decay of this band, as presently configured, is very fragmented; it populates six different states at 1409.3, 1808.8, 2300.0, 2523.5, 2540.2 and 2608.1 keV. These in turn, decay towards band $D2$ and the $15/2^-$ state at 1200.2 keV through 19 transitions. For many of the latter, it was possible to deduce angular-distribution coefficients and anisotropies, and this resulted in firm spin-parity assignments for all states. Consequently, the $Q4$ bandhead populated by the 423-keV transition is now fixed at 3370.8 keV with spin-parity $23/2^-$. In addition to the connecting transitions towards the newly observed low-lying states, the 418- and 511-keV transitions identified here for the first time link the $27/2^-$ and $31/2^-$ levels of band $Q4$ to the previously known $25/2^-$ and $29/2^-$ states of band $D6$, herewith providing further confidence to the spin-parity assignments.

Band $Q5$ was first reported in Ref.~\cite{karl1996} (as band 2), and linked to the $39/2^-$ state of band $Q4$ by the 882-keV $\gamma$ ray. While some of the previously reported transitions in this band are confirmed here, its structure and decay-out have been modified. This change was necessitated by the identification of: (i) the 477-keV line populating the $33/2^{(-)}$ bandhead at 5151.7 keV, (ii) a new 1311-keV transition placed in the middle of the band, (iii) three transitions of 702, 818 and 869 keV linking the sequence to band $Q4$ and (iv) the 1234-keV transition deexciting the bandhead of $Q5$ towards band $D3$. In addition, several new interconnecting transitions between bands $Q5$ and $Q6$ were identified in the $57/2 - 65/2$ spin range; these are attributed to the mixing between the two structures. The $E2$ character of the 1235-keV transition from band $Q6$ to the $61/2^{(-)}$ state of band $Q5$ fixes the parity of the two sequences as being the same. A weak 1486-keV $\gamma$ ray observed in coincidence with transitions in band $Q5$ is placed above the $69/2^{(-)}$ level and in parallel to the 15756.4-keV level depopulated by the 1500-keV $\gamma$ ray. The spins of band $Q5$ are fixed by the $\Delta I=1$ transitions connecting it to band $Q4$ where spin-parity assignments are firm. 
  
Similarly, band $Q6$ was first reported in Ref. \cite{karl1996} (as band 3), and is confirmed by the present results, with the exception of the highest 1711- and 1824-keV transitions which were not observed. The $65/2^{(-)}$ state decays towards the $61/2^{(-)}$ level of band $Q5$ through the 1235-keV $E2$ transition and towards an intermediate ($61/2^-$) state which in turn decays to the $57/2^{(-)}$ level of band $Q5$. These connecting transitions between bands $Q6$ and $Q5$ establish the same parity for the two bands. The $Q6$ bandhead, populated by the 979-keV transition is at 8746.5 keV and has spin-parity $49/2^{(-)}$.

Band $Q7$ was first reported in Ref. \cite{karl1996} (as band 4), with the suggestion that it decays towards low-lying levels in band $Q4$. This sequence is confirmed here, with the exception of the 1320-keV $\gamma$ ray which was not observed. While the 1385-keV transition is observed, it is now placed differently and a new transition of 1460 keV is placed on top of the band. Four new decay-out transitions towards band $Q4$ have also been identified. The $E2$ character of the 922-keV link with band $Q4$ establishes the negative parity of band $Q7$. The $Q7$ bandhead, populated by the 703-keV transition, is now at 5147.7 keV with spin-parity $35/2^{-}$.

The quadrupole sequence $Q8$ was first reported in Ref.~\cite{karl1996} (as band 5), and linked to band $Q7$ through the 1172-keV line. Only the 1223- and 1337-keV $\gamma$ rays are confirmed by the present data, but five new transitions now extend the sequence to excitation energy and spin of 22828.1 keV and $93/2$, respectively; the parity of the band members could not be firmly established.  The bandhead energy is determined as 13150.1 keV.

Band $Q9$ was first reported in Ref. \cite{karl1996} (as a cascade of four transitions below band 6), with the suggestion that it decays towards band $D6$. The present results confirm the previously observed $\gamma$ rays, but the in-band transitions of 1158 and 1241 keV are now placed in band $Q10$. One new transition of 860 keV was added at the bottom of the band, and six lines of 1150, 1244, 1340, 1436, 1453 and 1497 keV have been placed above the 10169.9-keV level depopulated by the 1061-keV transition reported previously~\cite{karl1996}. Band $Q9$ has been linked through the $758-898$-keV cascade to band $Q5$, and through the 775-keV transition and the $699 -1001$-keV cascade to band $Q11$. The $Q9$ bandhead is located at 7284.3 keV with spin-parity $43/2^{(-)}$.

Band $Q10$ was labeled band 6 in Ref. \cite{karl1996} where it was first suggested that it decays towards band $Q9$ through the 1201-keV line. This band is confirmed here but the highest $\gamma$ rays at 1779 and 1856 keV are not observed in the present data. The 1158- and 1241-keV transitions assigned previously to band $Q9$ are now placed in the bottom of band $Q10$. The $Q10$ bandhead  is at 11328.0 keV with spin-parity $59/2^{(-)}$.

Band $Q11$ is new to this work and is composed of four levels. It is fed by the 775-keV line from the $43/2^{(-)}$ state of band $Q9$. Its decay could not be established, but the feeding transitions from band $Q9$ fix the energy and spin-parity of the $Q11$ bandhead to 4040.1 keV and $27/2^{(-)}$.

\section{\label{sec-dis} Discussion} 
In this section, results of calculations performed within the framework of the cranked Nilsson-Strutinsky (CNS) model and of tilted axis cranking covariant density functional theory (TAC-CDFT) are presented. 
It should be noted that the present calculations focus mainly on the medium and high-spin structures where pairing effects are less important. Specific calculations for the low-spin part of the $^{133}$Ce nucleus have been performed in other formalisms such as the pair truncated shell model~\cite{PhysRevC.69.054309,PhysRevC.83.034321}, nucleon pair approximations~\cite{PhysRevC.76.054305} and other empirical models with pairing~\cite{ma1987}. 

\subsection{CNS Calculations}
The level structure of $^{133}$Ce, with 58 protons  and 75 neutrons, can be considered to arise from an interaction  
between eight valence proton particles above the $Z=50$ major shell  and seven neutron holes in the $N=82$ major shell.  In the low-energy regime, the nucleus is  expected to be characterized by a small deformation,  $\varepsilon_2 \sim 0.15-0.20$. Thus,  it is convenient to express the single-particle states in terms  of $j$-shell quantum numbers.  

In the CNS formalism \cite{ing-phys-rep,Ben85,Car06,Afa95}, the nucleus rotates about one of its principal axes and pairing is neglected. The deformation is optimized for each single-particle configuration under consideration.  The configurations are labelled by the number of particles in low-$j$ and high-$j$ orbitals, respectively, in the different ${\cal N}$-shells. The configurations can be defined relative to a $^{132}$Sn core as,
\[
\pi(dg)^{p_1}(h_{11/2})^{p_2} \nu(sd)^{-n_1}(h_{11/2})^{-n_2}(hf)^{n_3}(i_{13/2})^{n_4},
\]
for which we will use the short hand notation $[p_1 p_2,n_1 n_2 (n_3 n_4)]$.
The pseudospin partners $d_{5/2}g_{7/2}$ $(dg)$, $s_{1/2}d_{3/2}$ $(sd)$ and
$h_{9/2}f_{7/2}$ $(hf)$ are not distinguished in this formalism.
Note that all particles are listed; i.e., not only the particles considered as active (unpaired). 
Note also that the labels do not refer to
the pure $j$-shells, but rather to the dominating amplitudes in the Nilsson orbitals. In some cases, for an odd number of particles in a group, the signature will be specified as a  subscript $+ (\alpha = +1/2)$ or $- (\alpha = -1/2)$. The $A=130$ parameters introduced in Refs.~\cite{ing-phys-rep,Ben85} have been used for the $^{133}$Ce calculations.

\begin{figure}[!t]
\centering
\includegraphics[width=0.85\columnwidth]{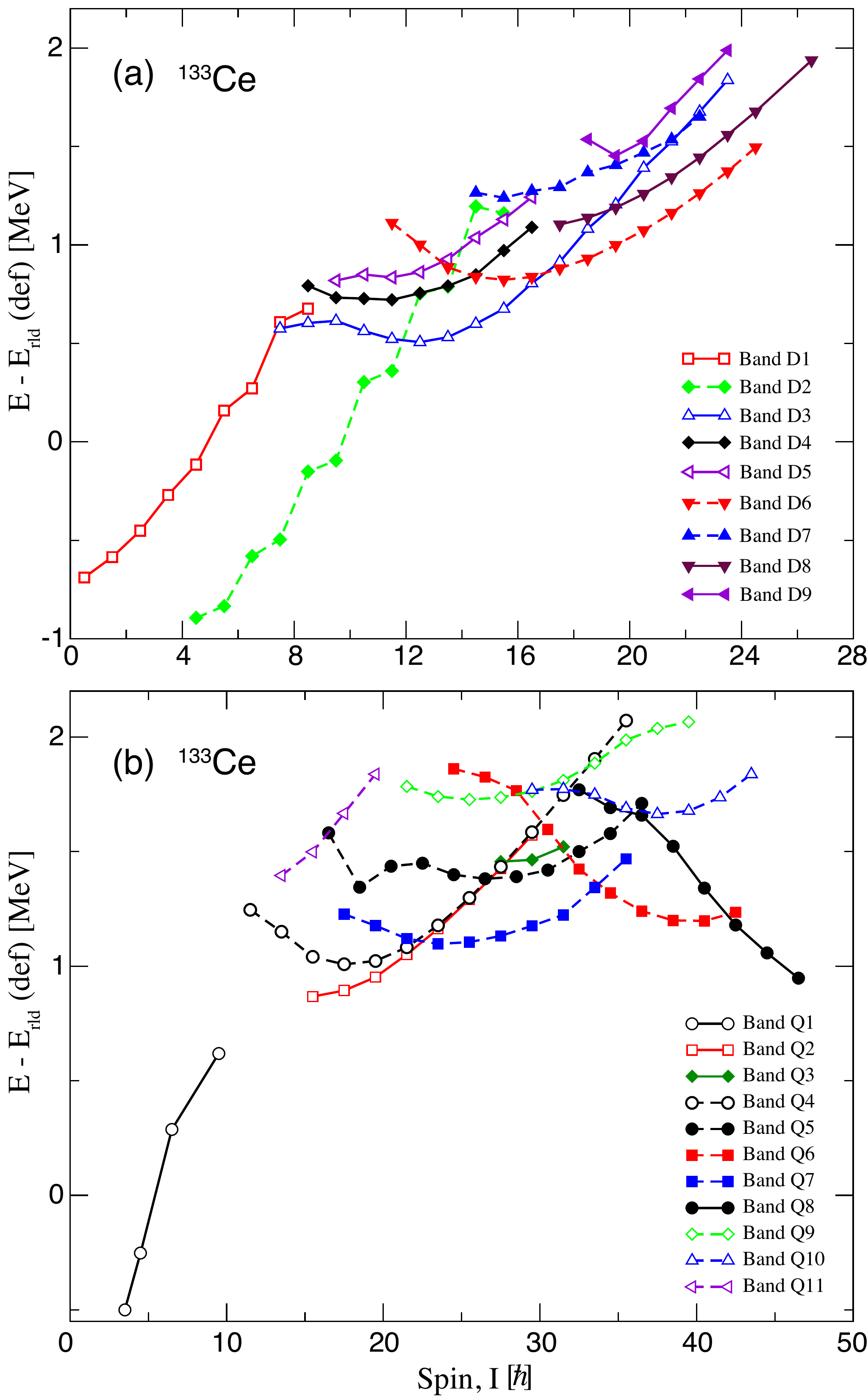}
\caption{\label{FIG7} (Color online) Energies relative to a standard rotating liquid drop reference calculated for the experimental bands observed in $^{133}$Ce. The Harris parameters used are: ${\cal J}_0 =22 \hbar^2$ MeV$^{-1}$ and  ${\cal J}_1 =11 \hbar^4$ MeV$^{-3}$.  }
\end{figure}

The  lowest proton configuration has eight protons in the $\pi g_{7/2}$ and $\pi d_{5/2}$ orbitals which interact and are strongly mixed. Higher angular momenta from proton configurations can be obtained by exciting one, two or three protons from these $\pi g_{7/2}$ and $\pi d_{5/2}$ orbitals into the $\pi h_{11/2}$ states. The lowest observed bands are characterized by configurations with one neutron hole in the $\nu d_{3/2}$ and $\nu s_{1/2}$ orbitals which also interact and mix strongly. Higher angular momenta can also be obtained from neutron configurations with one, two or three holes in the $\nu h_{11/2}$ orbital instead. Many more excited states and very-high angular momenta can result from neutron excitations above the $N=82$ shell gap into the  $\nu f_{7/2}, \nu h_{9/2}$ and $\nu i_{13/2}$ orbitals and proton excitations from the $\pi g_{9/2}$ state across the $Z=50$ shell gap. 

\begin{figure}[ht]
\centering\includegraphics[width=0.9\columnwidth]{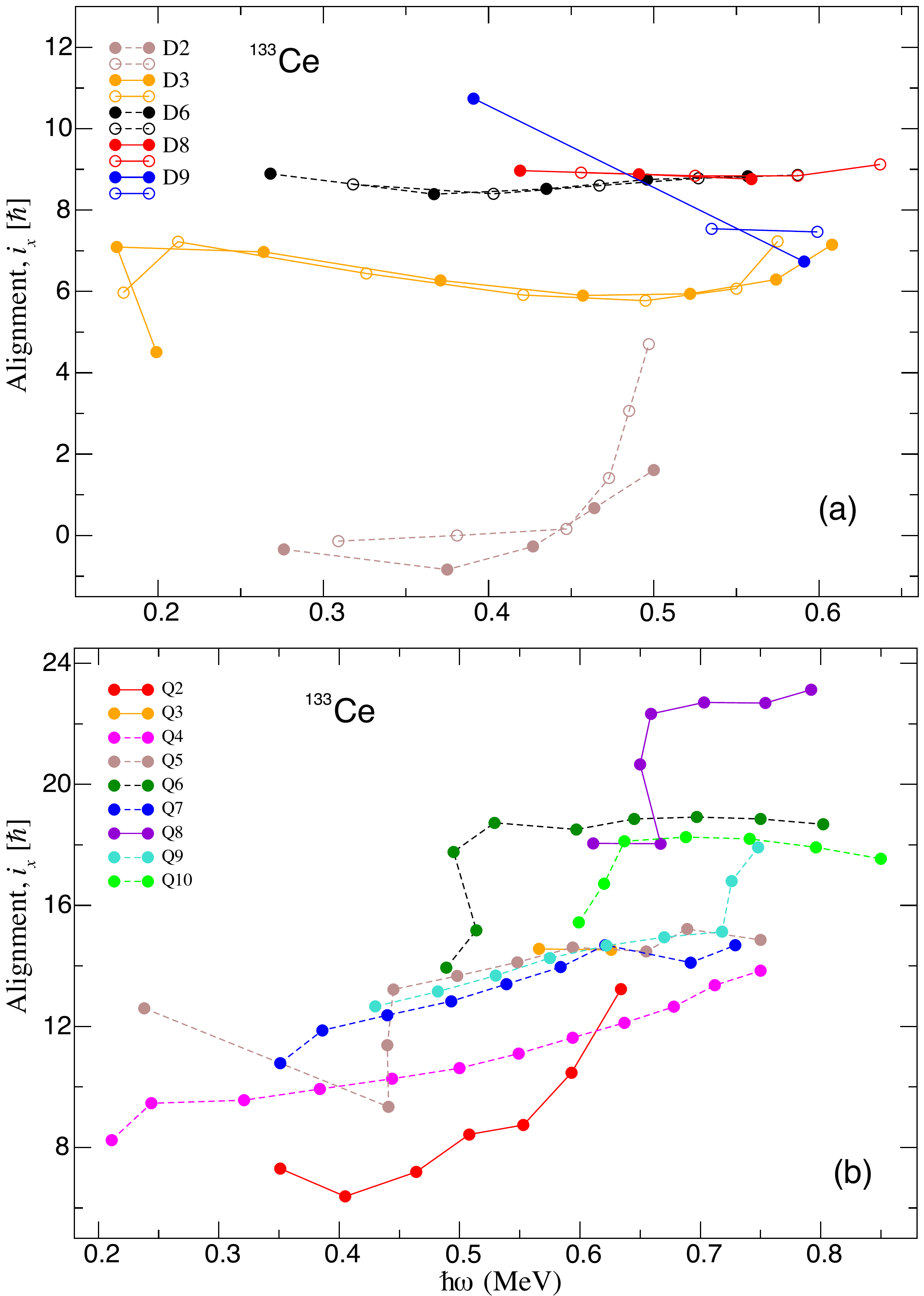}
\caption{\label{ix-d} (Color online) (a) Experimental alignments for bands $D2$, $D3$, $D6$, $D8$ and $D9$. (b) Experimental alignments for bands $Q2 - Q10$. The Harris parameters used to obtain these plots are: ${\cal J}_0 =22 \hbar^2$ MeV$^{-1}$ and  ${\cal J}_1 =11 \hbar^4$ MeV$^{-3}$.  }
\end{figure}

Energies relative to a standard rotating liquid drop reference for the experimental bands observed in $^{133}$Ce are presented in Fig.~\ref{FIG7}. As in the case of other triaxial bands observed in this mass region, the variation of the energy relative to a rotating liquid drop for medium- and high-spin bands has a parabolic behavior. In contrast, the lowest excited bands involving only one neutron hole have an upsloping behavior resulting from the increasing importance of pairing with decreasing spin, as recently discussed; $e. g.$, in Ref. \cite{138-high}. It should be noted that the dipole bands have a more pronounced curvature than the high-spin sequences. The configuration assignments discussed below are proposed by achieving the best possible agreement between the experimental and calculated energies and spins for each band. 
 
The experimental alignments for the $^{133}$Ce $D$ and $Q$ bands are plotted as a function of rotational frequency in Fig. \ref{ix-d}. They reveal the contribution of the active nucleons to the total angular momentum and have been used to guide the choice of associated CNS configurations.   
 
\subsubsection{The dipole bands }
\begin{figure*}[ht]
\centering\includegraphics[width=0.6\textwidth]{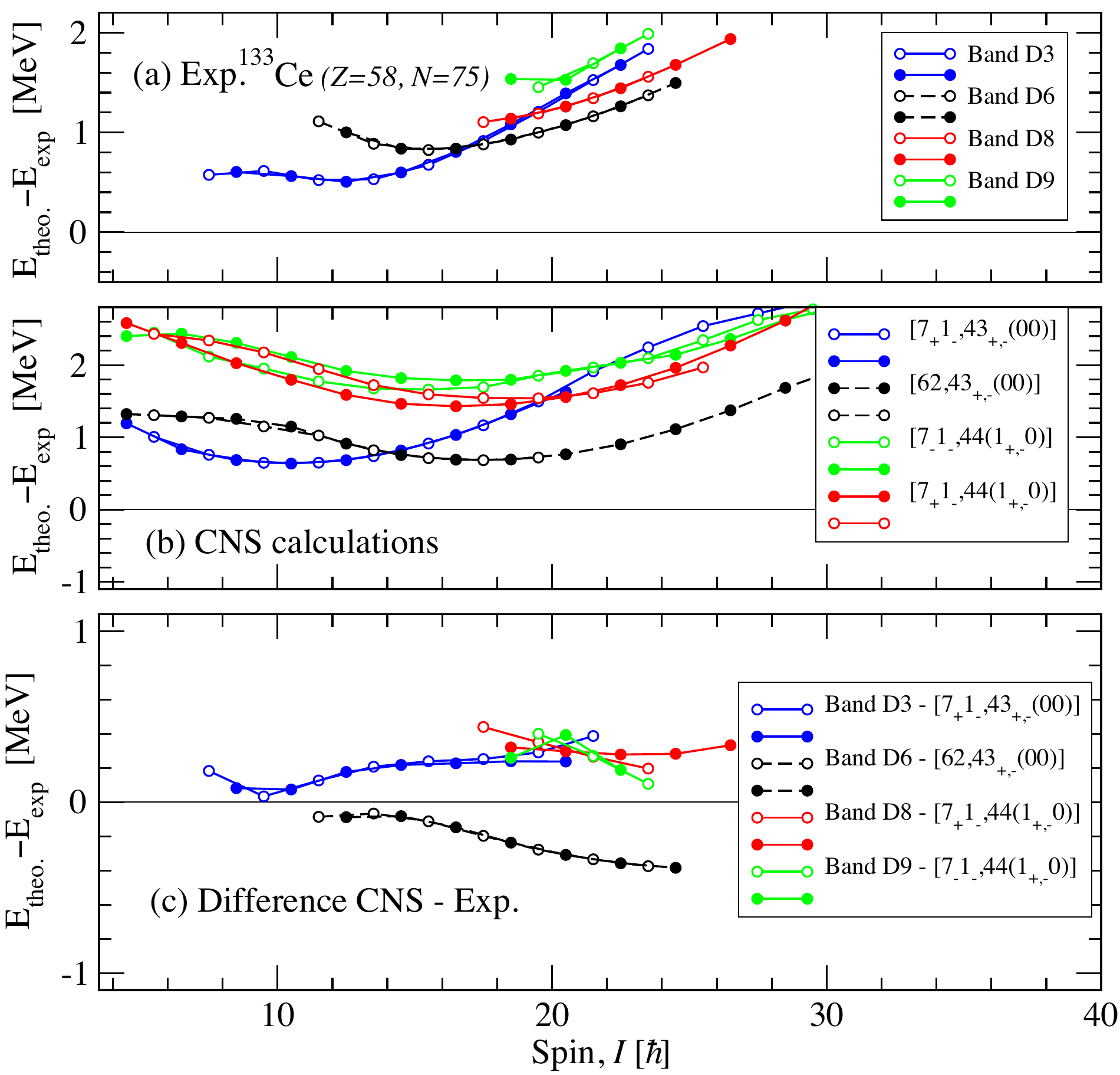}
\caption{(Color online) The observed bands $D3$, $D6$, $D8$ and $D9$ of $^{133}$Ce are
  shown relative to a rotating liquid drop reference in  the upper panel, with
  the calculated configurations assigned to these bands shown relative to the
  same reference in the middle panel. The lower panel shows the difference 
between calculations and experiment.}
\label{expth-D1}
\end{figure*}

The $D$ bands may be recognized in Fig.~\ref{FIG7}(a) as signature-doublet sequences. The lack of signature splitting signals an instability in a tilt of the rotational axis with respect to the principal axes and the need to carry out tilted axis cranking (TAC) calculations. However, as the  shape and energy of the bands are well accounted for by the CNS calculations (see Fig.~\ref{expth-D1}) with good accuracy, in this section, the interpretation will be based on the results of this model. 

A comparison between the experimental bands $D3$, $D6$, $D8$ and $D9$ and the results of the CNS calculations are provided in Fig. \ref{expth-D1}, where the proposed configurations are given as well.  The data and calculations agree to within $\pm 0.5$ MeV. The calculated configurations are all built on triaxial shapes, with quadrupole deformations decreasing slightly, from $\varepsilon_2 \approx 0.22$ to $\varepsilon_2 \approx 0.18$, with increasing spin, while the triaxiality parameter remains nearly constant with  $\gamma \approx +24^{\circ}$.

The configurations of bands $D3$ and $D6$ involve only orbitals from below the $N=82$ shell closure, while those of bands $D8$ and $D9$ are associated with one neutron in the $\nu (f_{7/2},h_{9/2})$ subshell. These $D3$ and $D6$ configurations are the same as those proposed previously~\cite{133ce-a,ma1987}.  The configurations assignments for the new bands $D8$ and $D9$, $[7_{\pm}1_-,44(1_{\pm}0)]$, or $\pi (d_{5/2},g_{7/2})h_{11/2}  \otimes \nu h_{11/2} (h_{9/2},f_{7/2})$ in terms of spherical orbitals, are similar to those assigned to bands with similar properties in neighboring nuclei; $e. g.$, bands 10 and 11 in $^{136}$Nd \cite{pet1996} and a newly identified sequence in $^{134}$Ce~\cite{pet2015}. These are associated with a neutron excitation  from the $\nu h_{11/2}$ orbital to the $\nu (h_{9/2},f_{7/2})$ state when compared to the $[7_{\pm} 1_-,43_{\pm}(00)]$ configuration of bands $D3$ and $D4$ towards which they decay.  The alignment of band $D8$ is similar to that of band $D6$ [see Fig.~\ref{ix-d}(a)], even though the proposed configurations are quite different. This can be explained as follows: although the $D8$ configuration has one fewer $h_{11/2}$ proton than that of band $D6$, it also has one neutron in the $(h_{9/2},f_{7/2})$ state, which contributes an amount of aligned angular momentum similar to that of the ``missing" $h_{11/2}$ proton.  
The irregular behavior of the aligned angular momentum exhibited by band $D9$ is not easy to understand -- it might possibly be due to the crossing/interaction  with one or more other unobserved, close-lying band(s).   
\subsubsection{The quadrupole high-spin bands }

\begin{figure*}[ht]
\centering\includegraphics[width=0.8\textwidth]{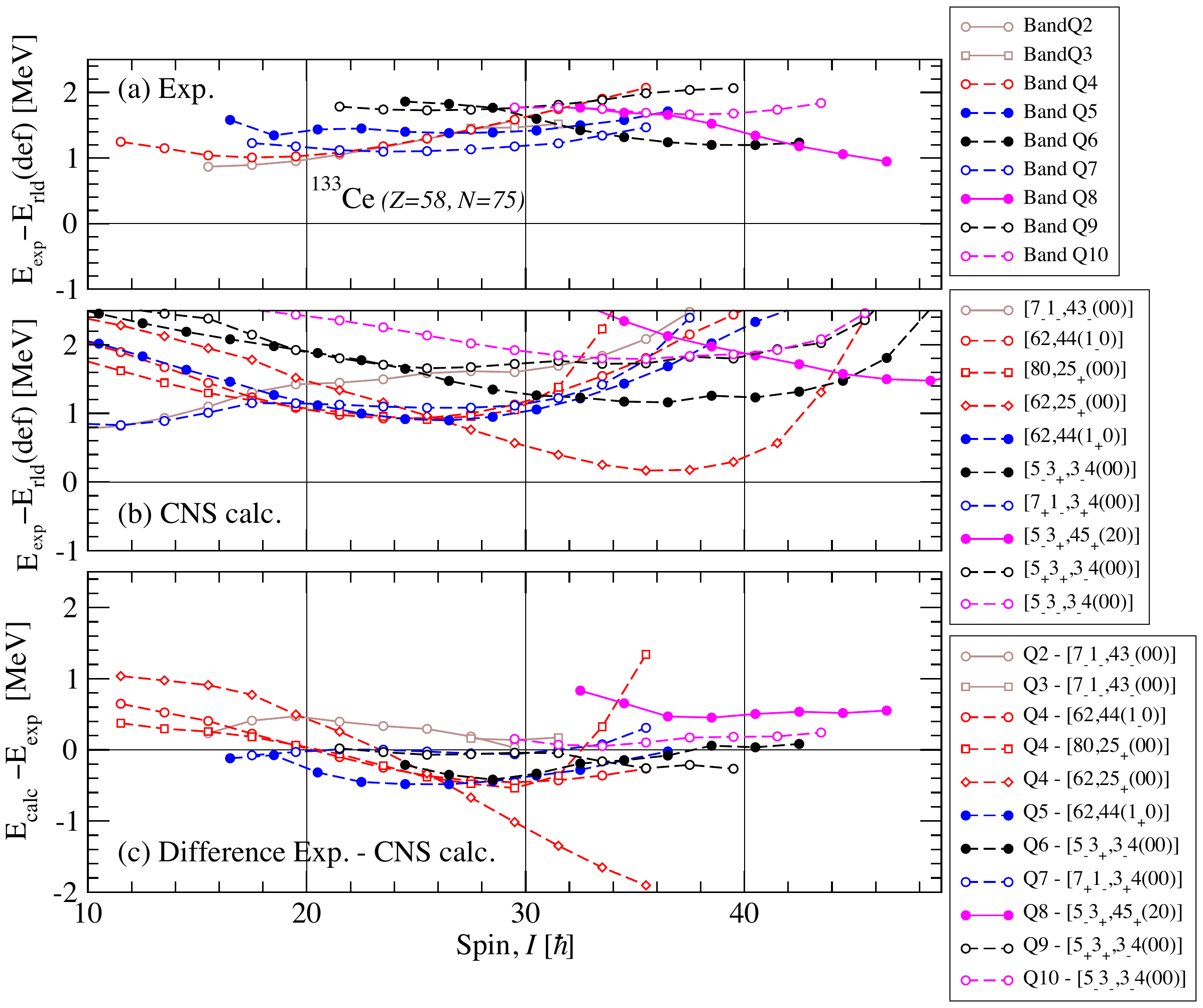}
\caption{(Color online) The observed bands $Q2$-$Q10$ of $^{133}$Ce are provided relative to a rotating liquid drop reference (upper panel), and the corresponding calculated configurations are displayed relative to the same reference (middle panel). The difference 
between calculations and experiment are provided at the bottom. The same $[7_- 1_-,43_-(00)]$ configuration  is assigned to bands $Q2$ and $Q3$, but with different triaxial deformation (see text). As the minima are very close in energy, only one curve is calculated and presented in the middle panel for this configuration.}
\label{expth-Q}
\end{figure*}

A similar comparison between experiment and theory for the $Q$-bands is presented in Fig.~\ref{expth-Q}, with the proposed configurations included here as well. 

Band $Q2$ (Fig.~\ref{FIG1}) has a minimum in the [$(E-E_{rld})$, $I$] plane [Fig.~\ref{expth-Q}(a)] at the lowest spin, and the experimental alignment [Fig.~\ref{ix-d} (b)] increases from 6 to 12 $\hbar$ over the spin range involved. The  $[7_- 1_-,43_-(00)]$ configuration is assigned, based on a comparison with the CNS calculations in Fig.~\ref{expth-Q}(b). This configuration represents a simple proton excitation relative to the $[7_+1_-,43_{\pm}(00)]$ assignment to band $D3$ towards which band $Q2$ decays (from the positive to the negative signature of the $h_{11/2}$ orbital). 
In the calculations, the deformation of the  $[7_- 1_-,43_-(00)]$ configuration changes abruptly from  $(\varepsilon_2 , \gamma) =( 0.18, +12^{\circ})$ at $I=35/2$ to $(\varepsilon_2 , \gamma) =( 0.17, -25^{\circ})$ at $I=39/2$. Since such an abrupt variation in the lower part of band $Q2$ is not observed, it is natural to suggest that its deformation  remains $(\varepsilon_2 , \gamma) =( 0.17, -25^{\circ})$.  

Over the observed spin range, the configuration of band $Q3$ should be similar to that of band $Q2$ to which it decays. As a result, it is assigned the same configuration as that of band $Q2$, $[7_- 1_-,43_-(00)]$, but with a different deformation;  $(\varepsilon_2 , \gamma) =( 0.16, -60^{\circ})$.  The change of the triaxial deformation between the bands $Q2$ and $Q3$ induces a smaller moment of inertia for $\gamma=-60^{\circ}$, and can account for the higher alignment observed for band $Q3$ [Fig.~\ref{ix-d} (b)].  

Band $Q4$ was originally discussed in Ref. \cite{karl1996} using total routhian surface (TRS) calculations, and it was concluded that the dominant configuration is $\nu (h_{11/2})^3$, with the gradual alignment of $s_{1/2}$ and $d_{3/2}$ neutrons being partially responsible for the smooth increase of the alignment. The present CNS calculations for the $[80,25_{+}(00)]$ configuration corresponding to $\nu (h_{11/2})^3$ exhibit a sharp upslope at high spins which is not seen experimentally (compare the top and middle panels of Fig. \ref{expth-Q}). This behavior practically excludes the  $[80,25_{+}(00)]$ configuration for the high-spin part of band $Q4$. A possible alternative would be $[62,25_+(00)]$, which includes an additional pair of $\pi h_{11/2}$ protons relative to the low-spin $[80,25_{+}(00)]$ configuration. The nature of band $Q4$ would, therefore, change from low to high spins through a crossing between the  $[80,25_{+}(00)]$ and $[62,25_+(00)]$ configurations. However, this $[62,25_+(00)]$ configuration is found to lie too low in energy at high spins (by $\sim$2 MeV) relative to the experimental band (see Fig. \ref{expth-Q}). In addition, it is sharply downsloping with increasing spin, in contrast with the observed behavior of band $Q4$ which is upsloping at high spins. The preferred CNS configuration for band $Q4$ is then $[62,44(1_-0)]$, which involves one neutron excited in the $(h_{9/2},f_{7/2})$ subshell above $N=82$. This suggested configuration provides enough angular momentum to induce a smooth increase of the energy at the highest observed spins.  As can be see in Fig. \ref{expth-Q}(c), the difference between the experimental $Q4$ energies and the $[62,44(1_-0)]$ configuration increases slightly at low spins where the pairing correlations, neglected in the present CNS calculations, become important.  The deformation of the  $[62,44(1_-0)]$ configuration is calculated to change gradually from  $(\varepsilon_2 , \gamma) =( 0.2, 0^{\circ})$ at $I=23/2$ to $(\varepsilon_2 , \gamma) =( 0.18, 28^{\circ})$ at $I=71/2$.

Bands $Q5$, $Q7$ and $Q9$ were also discussed in Ref. \cite{karl1996} (as bands  2, 4 and 7). However, the proposed assignments were made based on assumed excitation energies, spins and parities, resulting in a proposed near degeneracy for bands $Q5$ and $Q7$. In the present investigation, the energy and spins of these bands have now been firmly established, and this degeneracy is no longer present [Fig. \ref{FIG7}(b)]. Consequently, the suggested interpretation with bands  $Q5$ and $Q7$ as signature partners of the $\nu h_{11/2}^3 \otimes \pi h_{11/2}^2$ configuration is no longer valid. As evidenced by Fig.~\ref{ix-d}(b), the alignments of bands $Q5$, $Q7$ and $Q9$ are higher than that of band $Q4$, herewith suggesting similar orbital occupations for the three sequences. In fact, among the calculated, low-lying negative-parity configurations, the $[62,44(1_+0)]$,   $[7_+ 1_-,3_+4(00)]$ and $[5_+ 3_+,3_-4(00)]$ ones account well for the observed $Q5$, $Q7$ and $Q9$ respective behavior. 
The calculated deformation of the $[62, 44(1_+0)]$ configuration changes gradually from $(\varepsilon_2, \gamma) = (0.22, 11^\circ)$ at $I = 33/2$ to $(\varepsilon_2, \gamma) = (0.19, 29^\circ)$ at $I = 73/2$, that for the $[7_+1_-; 3_+4(00)]$ configuration changes from $(\varepsilon_2, \gamma) = (0.17,-20^\circ)$ at $I = 35/2$ to  $(\varepsilon_2, \gamma) = (0.10, -47^\circ)$ at $I = 71/2$, and that for the $[5_+3_+, 3_-4(00)]$ configuration changes from  $(\varepsilon_2, \gamma) = (0.20, 22^\circ)$ at $I = 43/2$ to  $(\varepsilon_2, \gamma) = (0.17,24^\circ$ at $I = 59/2$. The latter also jumps to a minimum with $(\varepsilon_2, \gamma) = (0.15,-30^\circ)$ up to spin $I=71/2$, and then at the highest observed spins, jumps back to the minimum with positive $\gamma$ $,(\varepsilon_2, \gamma) = (0.10, 28^\circ)$. The upbend observed at the highest spins in band $Q9$ can be explained by the jump between the two triaxial minima with positive and negative $\gamma$, or alternatively by a crossing with a configuration involving additional high-j orbitals.

Bands $Q6$ and $Q10$ have similar alignments, lying about $4 \hbar$ higher than those of bands $Q5$, $Q7$ and $Q9$. The CNS configurations with compatible excitation energies and spin-parity are $[5_-3_+,3_-4(00)]$ and  $[5_-3_-,3_-4(00)]$ for bands $Q6$ and $Q10$, respectively. These have similar number of occupied high-$j$ orbitals, with two additional $h_{11/2}$ particles relative to bands  $Q5$, $Q7$ and $Q9$. The calculated quadrupole deformation of the $[5_-3_+,3_-4(00)]$ configuration changes gradually from  $\varepsilon_2 =0.18$ at $I=53/2$ to $\varepsilon_2=0.08$ at $I=85/2$, while the triaxial deformation changes from $\gamma \approx +20^{\circ}$ ($I=61/2$), to $\gamma \approx -30^{\circ}$ ($I=65/2 - 73/2$), and to $\gamma \approx +20^{\circ}$ $I>73/2$, reflecting the softness of the shallow minimum with respect to $\gamma$ deformation. The calculated deformation of the  $[5_- 3_-,3_-4(00)]$ configuration changes gradually as well from  $(\varepsilon_2 , \gamma) =( 0.17, -32^{\circ})$ at $I=59/2$ to $(\varepsilon_2 , \gamma) =( 0.09, -47^{\circ})$ at $I=87/2$. 

Band $Q8$ develops only at high spins and at rather high excitation energy. Its downsloping behavior at the highest spins in the $(E-E_{rld})$-$vs$-$I$ plane [Fig.~\ref{FIG7}(b)] indicates the occupation of high-$j$ neutron orbitals from above the $N=82$ spherical shell closure. The lowest CNS configuration with a downsloping behavior in the spin range of interest is $[5_-3_+,45_+(20)]$, involving   two neutrons in the $(h_{9/2},f_{7/2})$ orbitals. The calculated deformation of the $[5_-3_+,45_+(20)]$ configuration changes gradually from  $(\varepsilon_2 , \gamma) =( 0.23, +17^{\circ})$ at $I=65/2$ to $(\varepsilon_2 , \gamma) =( 0.20, +30^{\circ})$ at $I=93/2$, thus higher than that of the other $Q$ bands. 

\subsection{TAC-CDFT Calculations}
As mentioned in the CNS calculations above, the dipole bands ($D$ bands) are interpreted as signature doublet sequences, and the observed small signature splitting may well correspond to rotational motion about an axis that does not coincide with any of the principal axes of the nucleus; i.e., the actual rotational axis may be tilted. Thus, for a detailed and comprehensive description of these bands, one may need to perform tilted axis cranking (TAC) calculations. To this end, the recently developed tilted axis cranking covariant density functional theory (TAC-CDFT), which enables a description of rotational excitations on the basis of a well-determined covariant density functional~\cite{Zhao2011Phys.Lett.B181,Meng2013FrontiersofPhysics55}, was employed. Due to computational limitations, the calculations were performed for band $D3$ only, as an example of the power of the approach. In principle, detailed systematic calculations for all the dipole bands would be warranted but very time-consuming; these will be presented in the future.

\begin{figure}[!b]
\includegraphics[width=8cm]{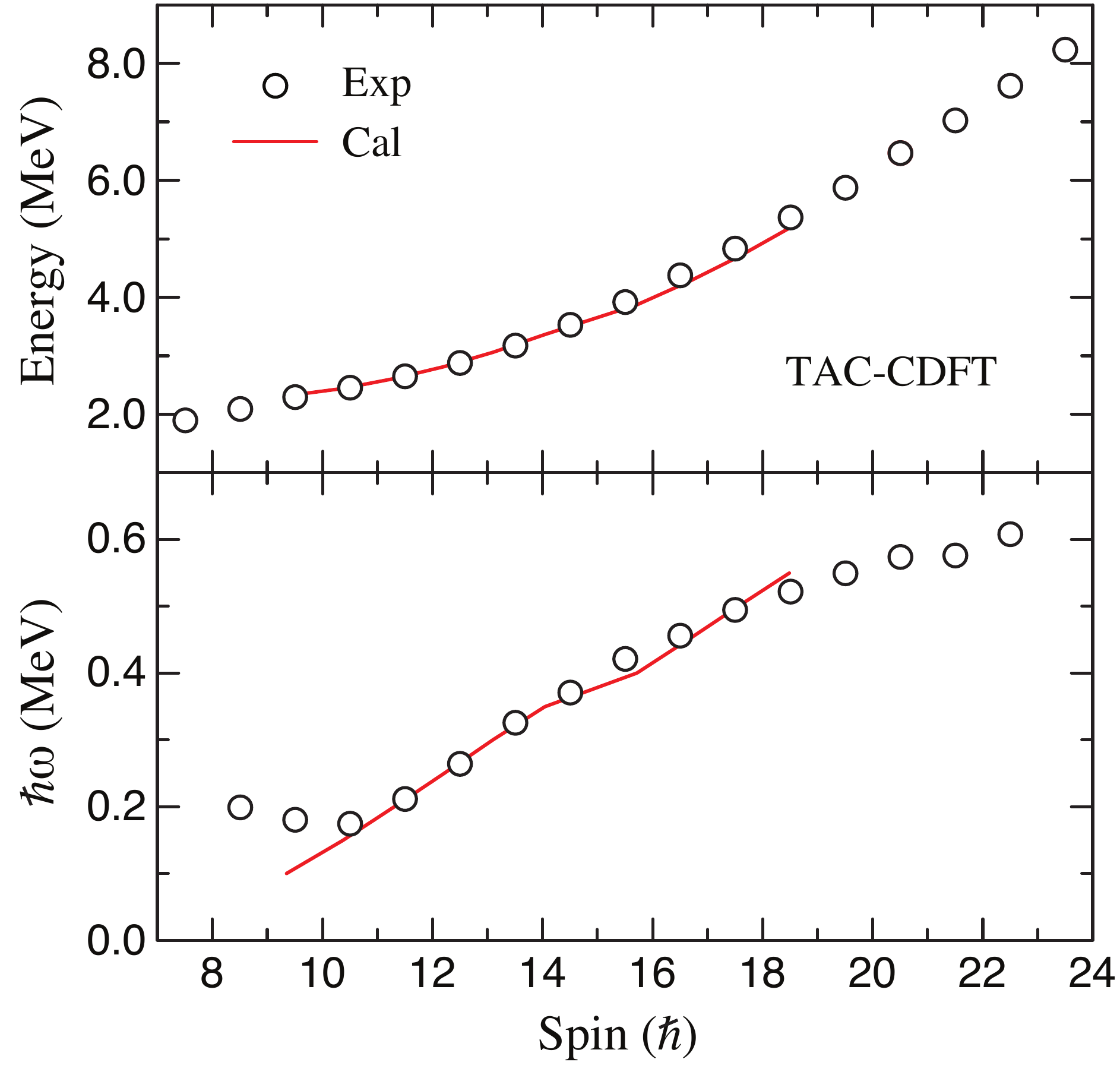}
\centering
\caption{(Color online) Calculated energy spectrum (upper panel) and frequency spectrum (lower panel) as a function of spin together with the data on band $D3$.}
\label{figpeng}
\end{figure}

For band $D3$, the point-coupling covariant density functional PC-PK1~\cite{Zhao2010Phys.Rev.C54319} was adopted in the particle-hole channel, while pairing correlations were neglected in the particle-particle channel. The Dirac equation for the nucleons is solved in a 3D cartesian harmonic oscillator basis with $N=10$ major shells. The self-consistent tilted axis cranking CDFT calculations were carried out based on the configuration $ \pi(h_{11/2})^{1}(2d_{5/2})^1(1g_{7/2})^{-2} \otimes \nu(h_{11/2})^{-3}(ds)^{-4} $. Note that, in this configuration, not all particles are considered to be active (unpaired), and the corresponding unpaired configuration should be written as $ \pi(h_{11/2})^{1}(2d_{5/2})^1 \otimes  \nu(h_{11/2})^{-1}$. 
The calculated energy spectrum as a function of spin with this configuration is found in the upper panel of Fig.~\ref{figpeng}: the calculations are in good agreement with the band $D3$ data. However, it is found that converged results are obtained mainly in the medium-spin part of the band, and the present configuration cannot be followed in either the very high-spin nor the low-spin part. The reason is illustrated in the lower panel of Fig.~\ref{figpeng}, where the experimental and calculated rotational frequency are compared as a function of the total angular momenta. The calculated total angular momenta agree well with the data from $I=23/2\hbar$ to $39/2\hbar$ and the increase is almost linear with frequency. This indicates that the moment of inertia is nearly constant and well reproduced by the present calculations. However, the data indicate changes with frequency around both $I=23/2$ and $I=39/2$ and this is expected to correspond to a change in the associated configuration around these angular momentum values, a feature which remains to be explored. 

It is worthwhile to mention that the calculated $B(M1)$ transition probabilities increase along the band, although the shears angle, the angle between the vectors of the proton and neutron angular momenta, drops from around $\sim 70^\circ$ to $\sim 35^\circ$. Obviously, this feature is at odds with the usual feature expected for a magnetic rotation band, where the $B(M1)$ values decrease as the shears angle close. Considering the fact that a substantial triaxial deformation $\gamma\sim20^\circ$ is obtained in the calculation, the existence of the chiral geometry and/or chiral vibration cannot be excluded for this configuration. To this end, an investigation based on three-dimensional cranking calculation would be most welcome. In addition, it was pointed out in the recent investigation for the yrast band in $^{135}\rm Nd$~\cite{Zhao2015PhysRevC.92.034319} that the transition from electric rotation to chiral vibration may occur, and that the pairing correlations play an important role in orienting the rotation axis. The study of a similar mechanism in the present  $^{133}\rm Ce$ nucleus would be interesting for future investigations.

\section{\label{sec-sum} Summary}
High-spin  states  in $^{133}$Ce have been populated with the $^{116}$Cd($^{22}$Ne,5{\it n}) reaction and the Gammasphere array.  A rather complete level scheme was developed, confirming most of the information reported previously, but adding new bands of quadrupole and dipole character and, thus extending the level scheme up to very high spins.  The observed bands were discussed using the CNS model and one dipole band was discussed within the framework of the TAC-CDFT approach. Possible configurations for the different bands were discussed. The global understanding of the observed bands adds strong support to the presence of pronounced triaxial deformation at medium and high spins, as well as signatures for rotation around different axes of the triaxial shape. The competition between assigned multiquasiparticle configurations are found to contribute to a rich diversity of collective phenomena in this nucleus.

\section{Acknowledgment }
CMP expresses his gratitude to Prof. I. Ragnarsson for providing the CNS codes, for the training on how to use them, and for enlightening comments on the theoretical interpretation of the results. This work has been supported in part by the  U. S. National Science Foundation Grants No. PHY07-58100, No. PHY-0822648, No. PHY-1068192, and No. PHY-1419765, and the U. S. Department of Energy, Office of Nuclear Physics, under Grant No. DE-FG02-94ER40834 (UM) and Contract No. DE-AC02-06CH11357 (ANL). This research used resources of ANL's ATLAS facility, which is a DOE Office of Science User Facility.

\bibliographystyle{apsrev4-1}

\bibliography{133Ce-Ayangeakaa}

\begin{thebibliography}{30}%
\makeatletter
\providecommand \@ifxundefined [1]{%
 \@ifx{#1\undefined}
}%
\providecommand \@ifnum [1]{%
 \ifnum #1\expandafter \@firstoftwo
 \else \expandafter \@secondoftwo
 \fi
}%
\providecommand \@ifx [1]{%
 \ifx #1\expandafter \@firstoftwo
 \else \expandafter \@secondoftwo
 \fi
}%
\providecommand \natexlab [1]{#1}%
\providecommand \enquote  [1]{``#1''}%
\providecommand \bibnamefont  [1]{#1}%
\providecommand \bibfnamefont [1]{#1}%
\providecommand \citenamefont [1]{#1}%
\providecommand \href@noop [0]{\@secondoftwo}%
\providecommand \href [0]{\begingroup \@sanitize@url \@href}%
\providecommand \@href[1]{\@@startlink{#1}\@@href}%
\providecommand \@@href[1]{\endgroup#1\@@endlink}%
\providecommand \@sanitize@url [0]{\catcode `\\12\catcode `\$12\catcode
  `\&12\catcode `\#12\catcode `\^12\catcode `\_12\catcode `\%12\relax}%
\providecommand \@@startlink[1]{}%
\providecommand \@@endlink[0]{}%
\providecommand \url  [0]{\begingroup\@sanitize@url \@url }%
\providecommand \@url [1]{\endgroup\@href {#1}{\urlprefix }}%
\providecommand \urlprefix  [0]{URL }%
\providecommand \Eprint [0]{\href }%
\providecommand \doibase [0]{http://dx.doi.org/}%
\providecommand \selectlanguage [0]{\@gobble}%
\providecommand \bibinfo  [0]{\@secondoftwo}%
\providecommand \bibfield  [0]{\@secondoftwo}%
\providecommand \translation [1]{[#1]}%
\providecommand \BibitemOpen [0]{}%
\providecommand \bibitemStop [0]{}%
\providecommand \bibitemNoStop [0]{.\EOS\space}%
\providecommand \EOS [0]{\spacefactor3000\relax}%
\providecommand \BibitemShut  [1]{\csname bibitem#1\endcsname}%
\let\auto@bib@innerbib\@empty
\bibitem [{\citenamefont {Ayangeakaa}\ \emph {et~al.}(2013)\citenamefont
  {Ayangeakaa}, \citenamefont {Garg}, \citenamefont {Anthony}, \citenamefont
  {Frauendorf}, \citenamefont {Matta}, \citenamefont {Nayak}, \citenamefont
  {Patel}, \citenamefont {Chen}, \citenamefont {Zhang}, \citenamefont {Zhao},
  \citenamefont {Qi}, \citenamefont {Meng}, \citenamefont {Janssens},
  \citenamefont {Carpenter}, \citenamefont {Chiara}, \citenamefont {Kondev},
  \citenamefont {Lauritsen}, \citenamefont {Seweryniak}, \citenamefont {Zhu},
  \citenamefont {Ghugre},\ and\ \citenamefont {Palit}}]{133ce-a}%
  \BibitemOpen
  \bibfield  {author} {\bibinfo {author} {\bibfnamefont {A.~D.}\ \bibnamefont
  {Ayangeakaa}}, \bibinfo {author} {\bibfnamefont {U.}~\bibnamefont {Garg}},
  \bibinfo {author} {\bibfnamefont {M.~D.}\ \bibnamefont {Anthony}}, \bibinfo
  {author} {\bibfnamefont {S.}~\bibnamefont {Frauendorf}}, \bibinfo {author}
  {\bibfnamefont {J.~T.}\ \bibnamefont {Matta}}, \bibinfo {author}
  {\bibfnamefont {B.~K.}\ \bibnamefont {Nayak}}, \bibinfo {author}
  {\bibfnamefont {D.}~\bibnamefont {Patel}}, \bibinfo {author} {\bibfnamefont
  {Q.~B.}\ \bibnamefont {Chen}}, \bibinfo {author} {\bibfnamefont {S.~Q.}\
  \bibnamefont {Zhang}}, \bibinfo {author} {\bibfnamefont {P.~W.}\ \bibnamefont
  {Zhao}}, \bibinfo {author} {\bibfnamefont {B.}~\bibnamefont {Qi}}, \bibinfo
  {author} {\bibfnamefont {J.}~\bibnamefont {Meng}}, \bibinfo {author}
  {\bibfnamefont {R.~V.~F.}\ \bibnamefont {Janssens}}, \bibinfo {author}
  {\bibfnamefont {M.~P.}\ \bibnamefont {Carpenter}}, \bibinfo {author}
  {\bibfnamefont {C.~J.}\ \bibnamefont {Chiara}}, \bibinfo {author}
  {\bibfnamefont {F.~G.}\ \bibnamefont {Kondev}}, \bibinfo {author}
  {\bibfnamefont {T.}~\bibnamefont {Lauritsen}}, \bibinfo {author}
  {\bibfnamefont {D.}~\bibnamefont {Seweryniak}}, \bibinfo {author}
  {\bibfnamefont {S.}~\bibnamefont {Zhu}}, \bibinfo {author} {\bibfnamefont
  {S.~S.}\ \bibnamefont {Ghugre}}, \ and\ \bibinfo {author} {\bibfnamefont
  {R.}~\bibnamefont {Palit}},\ }\href {\doibase 10.1103/PhysRevLett.110.172504}
  {\bibfield  {journal} {\bibinfo  {journal} {Phys. Rev. Lett.}\ }\textbf
  {\bibinfo {volume} {110}},\ \bibinfo {pages} {172504} (\bibinfo {year}
  {2013})}\BibitemShut {NoStop}%
\bibitem [{\citenamefont {Ma}\ \emph {et~al.}(1987)\citenamefont {Ma},
  \citenamefont {Paul}, \citenamefont {Beausang}, \citenamefont {Shi},
  \citenamefont {Xu},\ and\ \citenamefont {Fossan}}]{ma1987}%
  \BibitemOpen
  \bibfield  {author} {\bibinfo {author} {\bibfnamefont {R.}~\bibnamefont
  {Ma}}, \bibinfo {author} {\bibfnamefont {E.~S.}\ \bibnamefont {Paul}},
  \bibinfo {author} {\bibfnamefont {C.~W.}\ \bibnamefont {Beausang}}, \bibinfo
  {author} {\bibfnamefont {S.}~\bibnamefont {Shi}}, \bibinfo {author}
  {\bibfnamefont {N.}~\bibnamefont {Xu}}, \ and\ \bibinfo {author}
  {\bibfnamefont {D.~B.}\ \bibnamefont {Fossan}},\ }\href {\doibase
  10.1103/PhysRevC.36.2322} {\bibfield  {journal} {\bibinfo  {journal} {Phys.
  Rev. C}\ }\textbf {\bibinfo {volume} {36}},\ \bibinfo {pages} {2322}
  (\bibinfo {year} {1987})}\BibitemShut {NoStop}%
\bibitem [{\citenamefont {Emediato}\ \emph {et~al.}(1997)\citenamefont
  {Emediato}, \citenamefont {Rao}, \citenamefont {Medina}, \citenamefont
  {Seale}, \citenamefont {Botelho}, \citenamefont {Ribas}, \citenamefont
  {Oliveira}, \citenamefont {Cybulska}, \citenamefont {Espinoza-Qui\~nones},
  \citenamefont {Guimar\~aes}, \citenamefont {Rizzutto},\ and\ \citenamefont
  {Acquadro}}]{emediato1997}%
  \BibitemOpen
  \bibfield  {author} {\bibinfo {author} {\bibfnamefont {L.~G.~R.}\
  \bibnamefont {Emediato}}, \bibinfo {author} {\bibfnamefont {M.~N.}\
  \bibnamefont {Rao}}, \bibinfo {author} {\bibfnamefont {N.~H.}\ \bibnamefont
  {Medina}}, \bibinfo {author} {\bibfnamefont {W.~A.}\ \bibnamefont {Seale}},
  \bibinfo {author} {\bibfnamefont {S.}~\bibnamefont {Botelho}}, \bibinfo
  {author} {\bibfnamefont {R.~V.}\ \bibnamefont {Ribas}}, \bibinfo {author}
  {\bibfnamefont {J.~R.~B.}\ \bibnamefont {Oliveira}}, \bibinfo {author}
  {\bibfnamefont {E.~W.}\ \bibnamefont {Cybulska}}, \bibinfo {author}
  {\bibfnamefont {F.~R.}\ \bibnamefont {Espinoza-Qui\~nones}}, \bibinfo
  {author} {\bibfnamefont {V.}~\bibnamefont {Guimar\~aes}}, \bibinfo {author}
  {\bibfnamefont {M.~A.}\ \bibnamefont {Rizzutto}}, \ and\ \bibinfo {author}
  {\bibfnamefont {J.~C.}\ \bibnamefont {Acquadro}},\ }\href {\doibase
  10.1103/PhysRevC.55.2105} {\bibfield  {journal} {\bibinfo  {journal} {Phys.
  Rev. C}\ }\textbf {\bibinfo {volume} {55}},\ \bibinfo {pages} {2105}
  (\bibinfo {year} {1997})}\BibitemShut {NoStop}%
\bibitem [{\citenamefont {Hauschild}\ \emph {et~al.}(1995)\citenamefont
  {Hauschild}, \citenamefont {Wadsworth}, \citenamefont {Clark}, \citenamefont
  {Fallon}, \citenamefont {Fossan}, \citenamefont {Hibbert}, \citenamefont
  {Macchiavelli}, \citenamefont {Nolan}, \citenamefont {Schnare}, \citenamefont
  {Semple}, \citenamefont {Thorslund}, \citenamefont {Walker}, \citenamefont
  {Satula},\ and\ \citenamefont {Wyss}}]{karl1995}%
  \BibitemOpen
  \bibfield  {author} {\bibinfo {author} {\bibfnamefont {K.}~\bibnamefont
  {Hauschild}}, \bibinfo {author} {\bibfnamefont {R.}~\bibnamefont
  {Wadsworth}}, \bibinfo {author} {\bibfnamefont {R.}~\bibnamefont {Clark}},
  \bibinfo {author} {\bibfnamefont {P.}~\bibnamefont {Fallon}}, \bibinfo
  {author} {\bibfnamefont {D.}~\bibnamefont {Fossan}}, \bibinfo {author}
  {\bibfnamefont {I.}~\bibnamefont {Hibbert}}, \bibinfo {author} {\bibfnamefont
  {A.}~\bibnamefont {Macchiavelli}}, \bibinfo {author} {\bibfnamefont
  {P.}~\bibnamefont {Nolan}}, \bibinfo {author} {\bibfnamefont
  {H.}~\bibnamefont {Schnare}}, \bibinfo {author} {\bibfnamefont
  {A.}~\bibnamefont {Semple}}, \bibinfo {author} {\bibfnamefont
  {I.}~\bibnamefont {Thorslund}}, \bibinfo {author} {\bibfnamefont
  {L.}~\bibnamefont {Walker}}, \bibinfo {author} {\bibfnamefont
  {W.}~\bibnamefont {Satula}}, \ and\ \bibinfo {author} {\bibfnamefont
  {R.}~\bibnamefont {Wyss}},\ }\href {\doibase
  http://dx.doi.org/10.1016/0370-2693(95)00601-G} {\bibfield  {journal}
  {\bibinfo  {journal} {Phys. Lett. B}\ }\textbf {\bibinfo {volume} {353}},\
  \bibinfo {pages} {438 } (\bibinfo {year} {1995})}\BibitemShut {NoStop}%
\bibitem [{\citenamefont {Hauschild}\ \emph {et~al.}(1996)\citenamefont
  {Hauschild}, \citenamefont {Wadsworth}, \citenamefont {Clark}, \citenamefont
  {Hibbert}, \citenamefont {Fallon}, \citenamefont {Macchiavelli},
  \citenamefont {Fossan}, \citenamefont {Schnare}, \citenamefont {Thorslund},
  \citenamefont {Nolan}, \citenamefont {Semple},\ and\ \citenamefont
  {Walker}}]{karl1996}%
  \BibitemOpen
  \bibfield  {author} {\bibinfo {author} {\bibfnamefont {K.}~\bibnamefont
  {Hauschild}}, \bibinfo {author} {\bibfnamefont {R.}~\bibnamefont
  {Wadsworth}}, \bibinfo {author} {\bibfnamefont {R.~M.}\ \bibnamefont
  {Clark}}, \bibinfo {author} {\bibfnamefont {I.~M.}\ \bibnamefont {Hibbert}},
  \bibinfo {author} {\bibfnamefont {P.}~\bibnamefont {Fallon}}, \bibinfo
  {author} {\bibfnamefont {A.~O.}\ \bibnamefont {Macchiavelli}}, \bibinfo
  {author} {\bibfnamefont {D.~B.}\ \bibnamefont {Fossan}}, \bibinfo {author}
  {\bibfnamefont {H.}~\bibnamefont {Schnare}}, \bibinfo {author} {\bibfnamefont
  {I.}~\bibnamefont {Thorslund}}, \bibinfo {author} {\bibfnamefont {P.~J.}\
  \bibnamefont {Nolan}}, \bibinfo {author} {\bibfnamefont {A.~T.}\ \bibnamefont
  {Semple}}, \ and\ \bibinfo {author} {\bibfnamefont {L.}~\bibnamefont
  {Walker}},\ }\href {\doibase 10.1103/PhysRevC.54.613} {\bibfield  {journal}
  {\bibinfo  {journal} {Phys. Rev. C}\ }\textbf {\bibinfo {volume} {54}},\
  \bibinfo {pages} {613} (\bibinfo {year} {1996})}\BibitemShut {NoStop}%
\bibitem [{\citenamefont {Joss}\ \emph {et~al.}(1998)\citenamefont {Joss},
  \citenamefont {Paul}, \citenamefont {Clark}, \citenamefont {Lee},
  \citenamefont {Asztalos}, \citenamefont {Becker}, \citenamefont {Bernstein},
  \citenamefont {Cederwall}, \citenamefont {Deleplanque}, \citenamefont
  {Diamond}, \citenamefont {Fallon}, \citenamefont {Farris}, \citenamefont
  {Hauschild}, \citenamefont {Hibbert}, \citenamefont {Kelly}, \citenamefont
  {Macchiavelli}, \citenamefont {Nolan}, \citenamefont {O'Brien}, \citenamefont
  {Semple}, \citenamefont {Stephens},\ and\ \citenamefont
  {Wadsworth}}]{joss1998}%
  \BibitemOpen
  \bibfield  {author} {\bibinfo {author} {\bibfnamefont {D.~T.}\ \bibnamefont
  {Joss}}, \bibinfo {author} {\bibfnamefont {E.~S.}\ \bibnamefont {Paul}},
  \bibinfo {author} {\bibfnamefont {R.~M.}\ \bibnamefont {Clark}}, \bibinfo
  {author} {\bibfnamefont {I.~Y.}\ \bibnamefont {Lee}}, \bibinfo {author}
  {\bibfnamefont {S.~J.}\ \bibnamefont {Asztalos}}, \bibinfo {author}
  {\bibfnamefont {J.~A.}\ \bibnamefont {Becker}}, \bibinfo {author}
  {\bibfnamefont {L.}~\bibnamefont {Bernstein}}, \bibinfo {author}
  {\bibfnamefont {B.}~\bibnamefont {Cederwall}}, \bibinfo {author}
  {\bibfnamefont {M.~A.}\ \bibnamefont {Deleplanque}}, \bibinfo {author}
  {\bibfnamefont {R.~M.}\ \bibnamefont {Diamond}}, \bibinfo {author}
  {\bibfnamefont {P.}~\bibnamefont {Fallon}}, \bibinfo {author} {\bibfnamefont
  {L.~P.}\ \bibnamefont {Farris}}, \bibinfo {author} {\bibfnamefont
  {K.}~\bibnamefont {Hauschild}}, \bibinfo {author} {\bibfnamefont {I.~M.}\
  \bibnamefont {Hibbert}}, \bibinfo {author} {\bibfnamefont {W.~H.}\
  \bibnamefont {Kelly}}, \bibinfo {author} {\bibfnamefont {A.~O.}\ \bibnamefont
  {Macchiavelli}}, \bibinfo {author} {\bibfnamefont {P.~J.}\ \bibnamefont
  {Nolan}}, \bibinfo {author} {\bibfnamefont {N.~J.}\ \bibnamefont {O'Brien}},
  \bibinfo {author} {\bibfnamefont {A.~T.}\ \bibnamefont {Semple}}, \bibinfo
  {author} {\bibfnamefont {F.~S.}\ \bibnamefont {Stephens}}, \ and\ \bibinfo
  {author} {\bibfnamefont {R.}~\bibnamefont {Wadsworth}},\ }\href {\doibase
  10.1103/PhysRevC.58.3219} {\bibfield  {journal} {\bibinfo  {journal} {Phys.
  Rev. C}\ }\textbf {\bibinfo {volume} {58}},\ \bibinfo {pages} {3219}
  (\bibinfo {year} {1998})}\BibitemShut {NoStop}%
\bibitem [{\citenamefont {Afanasjev}\ \emph {et~al.}(1999)\citenamefont
  {Afanasjev}, \citenamefont {Fossan}, \citenamefont {Lane},\ and\
  \citenamefont {Ragnarsson}}]{ing-phys-rep}%
  \BibitemOpen
  \bibfield  {author} {\bibinfo {author} {\bibfnamefont {A.}~\bibnamefont
  {Afanasjev}}, \bibinfo {author} {\bibfnamefont {D.}~\bibnamefont {Fossan}},
  \bibinfo {author} {\bibfnamefont {G.}~\bibnamefont {Lane}}, \ and\ \bibinfo
  {author} {\bibfnamefont {I.}~\bibnamefont {Ragnarsson}},\ }\href {\doibase
  http://dx.doi.org/10.1016/S0370-1573(99)00035-6} {\bibfield  {journal}
  {\bibinfo  {journal} {Phys. Rep.}\ }\textbf {\bibinfo {volume} {322}},\
  \bibinfo {pages} {1 } (\bibinfo {year} {1999})}\BibitemShut {NoStop}%
\bibitem [{\citenamefont {Bengtsson}\ and\ \citenamefont
  {Ragnarsson}(1985)}]{Ben85}%
  \BibitemOpen
  \bibfield  {author} {\bibinfo {author} {\bibfnamefont {T.}~\bibnamefont
  {Bengtsson}}\ and\ \bibinfo {author} {\bibfnamefont {I.}~\bibnamefont
  {Ragnarsson}},\ }\href {\doibase
  http://dx.doi.org/10.1016/0375-9474(85)90541-X} {\bibfield  {journal}
  {\bibinfo  {journal} {Nucl. Phys. A}\ }\textbf {\bibinfo {volume} {436}},\
  \bibinfo {pages} {14 } (\bibinfo {year} {1985})}\BibitemShut {NoStop}%
\bibitem [{\citenamefont {Afanasjev}\ and\ \citenamefont
  {Ragnarsson}(1995)}]{Afa95}%
  \BibitemOpen
  \bibfield  {author} {\bibinfo {author} {\bibfnamefont {A.}~\bibnamefont
  {Afanasjev}}\ and\ \bibinfo {author} {\bibfnamefont {I.}~\bibnamefont
  {Ragnarsson}},\ }\href {\doibase
  http://dx.doi.org/10.1016/0375-9474(95)00196-8} {\bibfield  {journal}
  {\bibinfo  {journal} {Nucl. Phys. A}\ }\textbf {\bibinfo {volume} {591}},\
  \bibinfo {pages} {387 } (\bibinfo {year} {1995})}\BibitemShut {NoStop}%
\bibitem [{\citenamefont {Carlsson}\ and\ \citenamefont
  {Ragnarsson}(2006)}]{Car06}%
  \BibitemOpen
  \bibfield  {author} {\bibinfo {author} {\bibfnamefont {B.~G.}\ \bibnamefont
  {Carlsson}}\ and\ \bibinfo {author} {\bibfnamefont {I.}~\bibnamefont
  {Ragnarsson}},\ }\href {\doibase 10.1103/PhysRevC.74.011302} {\bibfield
  {journal} {\bibinfo  {journal} {Phys. Rev. C}\ }\textbf {\bibinfo {volume}
  {74}},\ \bibinfo {pages} {011302} (\bibinfo {year} {2006})}\BibitemShut
  {NoStop}%
\bibitem [{\citenamefont {Zhao}\ \emph {et~al.}(2011)\citenamefont {Zhao},
  \citenamefont {Zhang}, \citenamefont {Peng}, \citenamefont {Liang},
  \citenamefont {Ring},\ and\ \citenamefont {Meng}}]{Zhao2011Phys.Lett.B181}%
  \BibitemOpen
  \bibfield  {author} {\bibinfo {author} {\bibfnamefont {P.~W.}\ \bibnamefont
  {Zhao}}, \bibinfo {author} {\bibfnamefont {S.~Q.}\ \bibnamefont {Zhang}},
  \bibinfo {author} {\bibfnamefont {J.}~\bibnamefont {Peng}}, \bibinfo {author}
  {\bibfnamefont {H.~Z.}\ \bibnamefont {Liang}}, \bibinfo {author}
  {\bibfnamefont {P.}~\bibnamefont {Ring}}, \ and\ \bibinfo {author}
  {\bibfnamefont {J.}~\bibnamefont {Meng}},\ }\href {\doibase
  10.1016/j.physletb.2011.03.068} {\bibfield  {journal} {\bibinfo  {journal}
  {Phys. Lett. B}\ }\textbf {\bibinfo {volume} {699}},\ \bibinfo {pages} {181}
  (\bibinfo {year} {2011})}\BibitemShut {NoStop}%
\bibitem [{\citenamefont {Meng}\ \emph {et~al.}(2013)\citenamefont {Meng},
  \citenamefont {Peng}, \citenamefont {Zhang},\ and\ \citenamefont
  {Zhao}}]{Meng2013FrontiersofPhysics55}%
  \BibitemOpen
  \bibfield  {author} {\bibinfo {author} {\bibfnamefont {J.}~\bibnamefont
  {Meng}}, \bibinfo {author} {\bibfnamefont {J.}~\bibnamefont {Peng}}, \bibinfo
  {author} {\bibfnamefont {S.-Q.}\ \bibnamefont {Zhang}}, \ and\ \bibinfo
  {author} {\bibfnamefont {P.-W.}\ \bibnamefont {Zhao}},\ }\href {\doibase
  10.1007/s11467-013-0287-y} {\bibfield  {journal} {\bibinfo  {journal} {Front.
  Phys.}\ }\textbf {\bibinfo {volume} {8}},\ \bibinfo {pages} {55} (\bibinfo
  {year} {2013})}\BibitemShut {NoStop}%
\bibitem [{\citenamefont {Petrache}\ \emph {et~al.}(2012)\citenamefont
  {Petrache}, \citenamefont {Frauendorf}, \citenamefont {Matsuzaki},
  \citenamefont {Leguillon}, \citenamefont {Zerrouki}, \citenamefont {Lunardi},
  \citenamefont {Bazzacco}, \citenamefont {Ur}, \citenamefont {Farnea},
  \citenamefont {Rossi~Alvarez}, \citenamefont {Venturelli},\ and\
  \citenamefont {de~Angelis}}]{138-low}%
  \BibitemOpen
  \bibfield  {author} {\bibinfo {author} {\bibfnamefont {C.~M.}\ \bibnamefont
  {Petrache}}, \bibinfo {author} {\bibfnamefont {S.}~\bibnamefont
  {Frauendorf}}, \bibinfo {author} {\bibfnamefont {M.}~\bibnamefont
  {Matsuzaki}}, \bibinfo {author} {\bibfnamefont {R.}~\bibnamefont
  {Leguillon}}, \bibinfo {author} {\bibfnamefont {T.}~\bibnamefont {Zerrouki}},
  \bibinfo {author} {\bibfnamefont {S.}~\bibnamefont {Lunardi}}, \bibinfo
  {author} {\bibfnamefont {D.}~\bibnamefont {Bazzacco}}, \bibinfo {author}
  {\bibfnamefont {C.~A.}\ \bibnamefont {Ur}}, \bibinfo {author} {\bibfnamefont
  {E.}~\bibnamefont {Farnea}}, \bibinfo {author} {\bibfnamefont
  {C.}~\bibnamefont {Rossi~Alvarez}}, \bibinfo {author} {\bibfnamefont
  {R.}~\bibnamefont {Venturelli}}, \ and\ \bibinfo {author} {\bibfnamefont
  {G.}~\bibnamefont {de~Angelis}},\ }\href {\doibase
  10.1103/PhysRevC.86.044321} {\bibfield  {journal} {\bibinfo  {journal} {Phys.
  Rev. C}\ }\textbf {\bibinfo {volume} {86}},\ \bibinfo {pages} {044321}
  (\bibinfo {year} {2012})}\BibitemShut {NoStop}%
\bibitem [{\citenamefont {Petrache}\ \emph {et~al.}(2013)\citenamefont
  {Petrache}, \citenamefont {Ragnarsson}, \citenamefont {Ma}, \citenamefont
  {Leguillon}, \citenamefont {Konstantinopoulos}, \citenamefont {Zerrouki},
  \citenamefont {Bazzacco},\ and\ \citenamefont {Lunardi}}]{138-switch}%
  \BibitemOpen
  \bibfield  {author} {\bibinfo {author} {\bibfnamefont {C.~M.}\ \bibnamefont
  {Petrache}}, \bibinfo {author} {\bibfnamefont {I.}~\bibnamefont
  {Ragnarsson}}, \bibinfo {author} {\bibfnamefont {H.-L.}\ \bibnamefont {Ma}},
  \bibinfo {author} {\bibfnamefont {R.}~\bibnamefont {Leguillon}}, \bibinfo
  {author} {\bibfnamefont {T.}~\bibnamefont {Konstantinopoulos}}, \bibinfo
  {author} {\bibfnamefont {T.}~\bibnamefont {Zerrouki}}, \bibinfo {author}
  {\bibfnamefont {D.}~\bibnamefont {Bazzacco}}, \ and\ \bibinfo {author}
  {\bibfnamefont {S.}~\bibnamefont {Lunardi}},\ }\href {\doibase
  10.1103/PhysRevC.88.051303} {\bibfield  {journal} {\bibinfo  {journal} {Phys.
  Rev. C}\ }\textbf {\bibinfo {volume} {88}},\ \bibinfo {pages} {051303}
  (\bibinfo {year} {2013})}\BibitemShut {NoStop}%
\bibitem [{\citenamefont {Petrache}\ \emph {et~al.}(2015)\citenamefont
  {Petrache}, \citenamefont {Ragnarsson}, \citenamefont {Ma}, \citenamefont
  {Leguillon}, \citenamefont {Zerrouki}, \citenamefont {Bazzacco},\ and\
  \citenamefont {Lunardi}}]{138-high}%
  \BibitemOpen
  \bibfield  {author} {\bibinfo {author} {\bibfnamefont {C.~M.}\ \bibnamefont
  {Petrache}}, \bibinfo {author} {\bibfnamefont {I.}~\bibnamefont
  {Ragnarsson}}, \bibinfo {author} {\bibfnamefont {H.-L.}\ \bibnamefont {Ma}},
  \bibinfo {author} {\bibfnamefont {R.}~\bibnamefont {Leguillon}}, \bibinfo
  {author} {\bibfnamefont {T.}~\bibnamefont {Zerrouki}}, \bibinfo {author}
  {\bibfnamefont {D.}~\bibnamefont {Bazzacco}}, \ and\ \bibinfo {author}
  {\bibfnamefont {S.}~\bibnamefont {Lunardi}},\ }\href {\doibase
  10.1103/PhysRevC.91.024302} {\bibfield  {journal} {\bibinfo  {journal} {Phys.
  Rev. C}\ }\textbf {\bibinfo {volume} {91}},\ \bibinfo {pages} {024302}
  (\bibinfo {year} {2015})}\BibitemShut {NoStop}%
\bibitem [{\citenamefont {Leguillon}\ \emph {et~al.}(2013)\citenamefont
  {Leguillon}, \citenamefont {Petrache}, \citenamefont {Zerrouki},
  \citenamefont {Konstantinopoulos}, \citenamefont {Hauschild}, \citenamefont
  {Korichi}, \citenamefont {Lopez-Martens}, \citenamefont {Frauendorf},
  \citenamefont {Ragnarsson}, \citenamefont {Greenlees}, \citenamefont
  {Jakobsson}, \citenamefont {Jones}, \citenamefont {Julin}, \citenamefont
  {Juutinen}, \citenamefont {Ketelhut}, \citenamefont {Leino}, \citenamefont
  {Nieminen}, \citenamefont {Nyman}, \citenamefont {Peura}, \citenamefont
  {Rahkila}, \citenamefont {Ruotsalainen}, \citenamefont {Sandzelius},
  \citenamefont {Saren}, \citenamefont {Scholey}, \citenamefont {Sorri},
  \citenamefont {Uusitalo}, \citenamefont {H\"ubel}, \citenamefont
  {Neu\ss{}er-Neffgen}, \citenamefont {Al-Khatib}, \citenamefont {B\"urger},
  \citenamefont {Nenoff}, \citenamefont {Singh}, \citenamefont {Curien},
  \citenamefont {Hagemann}, \citenamefont {Herskind}, \citenamefont {Sletten},
  \citenamefont {Fallon}, \citenamefont {G\"orgen}, \citenamefont
  {Bednarczyk},\ and\ \citenamefont {Cullen}}]{140-high}%
  \BibitemOpen
  \bibfield  {author} {\bibinfo {author} {\bibfnamefont {R.}~\bibnamefont
  {Leguillon}}, \bibinfo {author} {\bibfnamefont {C.~M.}\ \bibnamefont
  {Petrache}}, \bibinfo {author} {\bibfnamefont {T.}~\bibnamefont {Zerrouki}},
  \bibinfo {author} {\bibfnamefont {T.}~\bibnamefont {Konstantinopoulos}},
  \bibinfo {author} {\bibfnamefont {K.}~\bibnamefont {Hauschild}}, \bibinfo
  {author} {\bibfnamefont {A.}~\bibnamefont {Korichi}}, \bibinfo {author}
  {\bibfnamefont {A.}~\bibnamefont {Lopez-Martens}}, \bibinfo {author}
  {\bibfnamefont {S.}~\bibnamefont {Frauendorf}}, \bibinfo {author}
  {\bibfnamefont {I.}~\bibnamefont {Ragnarsson}}, \bibinfo {author}
  {\bibfnamefont {P.~T.}\ \bibnamefont {Greenlees}}, \bibinfo {author}
  {\bibfnamefont {U.}~\bibnamefont {Jakobsson}}, \bibinfo {author}
  {\bibfnamefont {P.}~\bibnamefont {Jones}}, \bibinfo {author} {\bibfnamefont
  {R.}~\bibnamefont {Julin}}, \bibinfo {author} {\bibfnamefont
  {S.}~\bibnamefont {Juutinen}}, \bibinfo {author} {\bibfnamefont
  {S.}~\bibnamefont {Ketelhut}}, \bibinfo {author} {\bibfnamefont
  {M.}~\bibnamefont {Leino}}, \bibinfo {author} {\bibfnamefont
  {P.}~\bibnamefont {Nieminen}}, \bibinfo {author} {\bibfnamefont
  {M.}~\bibnamefont {Nyman}}, \bibinfo {author} {\bibfnamefont
  {P.}~\bibnamefont {Peura}}, \bibinfo {author} {\bibfnamefont
  {P.}~\bibnamefont {Rahkila}}, \bibinfo {author} {\bibfnamefont
  {P.}~\bibnamefont {Ruotsalainen}}, \bibinfo {author} {\bibfnamefont
  {M.}~\bibnamefont {Sandzelius}}, \bibinfo {author} {\bibfnamefont
  {J.}~\bibnamefont {Saren}}, \bibinfo {author} {\bibfnamefont
  {C.}~\bibnamefont {Scholey}}, \bibinfo {author} {\bibfnamefont
  {J.}~\bibnamefont {Sorri}}, \bibinfo {author} {\bibfnamefont
  {J.}~\bibnamefont {Uusitalo}}, \bibinfo {author} {\bibfnamefont
  {H.}~\bibnamefont {H\"ubel}}, \bibinfo {author} {\bibfnamefont
  {A.}~\bibnamefont {Neu\ss{}er-Neffgen}}, \bibinfo {author} {\bibfnamefont
  {A.}~\bibnamefont {Al-Khatib}}, \bibinfo {author} {\bibfnamefont
  {A.}~\bibnamefont {B\"urger}}, \bibinfo {author} {\bibfnamefont
  {N.}~\bibnamefont {Nenoff}}, \bibinfo {author} {\bibfnamefont {A.~K.}\
  \bibnamefont {Singh}}, \bibinfo {author} {\bibfnamefont {D.}~\bibnamefont
  {Curien}}, \bibinfo {author} {\bibfnamefont {G.~B.}\ \bibnamefont
  {Hagemann}}, \bibinfo {author} {\bibfnamefont {B.}~\bibnamefont {Herskind}},
  \bibinfo {author} {\bibfnamefont {G.}~\bibnamefont {Sletten}}, \bibinfo
  {author} {\bibfnamefont {P.}~\bibnamefont {Fallon}}, \bibinfo {author}
  {\bibfnamefont {A.}~\bibnamefont {G\"orgen}}, \bibinfo {author}
  {\bibfnamefont {P.}~\bibnamefont {Bednarczyk}}, \ and\ \bibinfo {author}
  {\bibfnamefont {D.~M.}\ \bibnamefont {Cullen}},\ }\href {\doibase
  10.1103/PhysRevC.88.014323} {\bibfield  {journal} {\bibinfo  {journal} {Phys.
  Rev. C}\ }\textbf {\bibinfo {volume} {88}},\ \bibinfo {pages} {014323}
  (\bibinfo {year} {2013})}\BibitemShut {NoStop}%
\bibitem [{\citenamefont {Zerrouki}\ \emph {et~al.}(2015)\citenamefont
  {Zerrouki}, \citenamefont {Petrache}, \citenamefont {Leguillon},
  \citenamefont {Hauschild}, \citenamefont {Korichi}, \citenamefont
  {Lopez-Martens}, \citenamefont {Frauendorf}, \citenamefont {Ragnarsson},
  \citenamefont {H{\"u}bel}, \citenamefont {Neu{\ss}er-Neffgen}, \citenamefont
  {Al-Khatib}, \citenamefont {Bringel}, \citenamefont {B{\"u}rger},
  \citenamefont {Nenoff}, \citenamefont {Sch{\"o}nwa{\ss}er}, \citenamefont
  {Singh}, \citenamefont {Curien}, \citenamefont {Hagemann}, \citenamefont
  {Herskind}, \citenamefont {Sletten}, \citenamefont {Fallon}, \citenamefont
  {G{\"o}rgen},\ and\ \citenamefont {Bednarczyk}}]{141nd}%
  \BibitemOpen
  \bibfield  {author} {\bibinfo {author} {\bibfnamefont {T.}~\bibnamefont
  {Zerrouki}}, \bibinfo {author} {\bibfnamefont {C.}~\bibnamefont {Petrache}},
  \bibinfo {author} {\bibfnamefont {R.}~\bibnamefont {Leguillon}}, \bibinfo
  {author} {\bibfnamefont {K.}~\bibnamefont {Hauschild}}, \bibinfo {author}
  {\bibfnamefont {A.}~\bibnamefont {Korichi}}, \bibinfo {author} {\bibfnamefont
  {A.}~\bibnamefont {Lopez-Martens}}, \bibinfo {author} {\bibfnamefont
  {S.}~\bibnamefont {Frauendorf}}, \bibinfo {author} {\bibfnamefont
  {I.}~\bibnamefont {Ragnarsson}}, \bibinfo {author} {\bibfnamefont
  {H.}~\bibnamefont {H{\"u}bel}}, \bibinfo {author} {\bibfnamefont
  {A.}~\bibnamefont {Neu{\ss}er-Neffgen}}, \bibinfo {author} {\bibfnamefont
  {A.}~\bibnamefont {Al-Khatib}}, \bibinfo {author} {\bibfnamefont
  {P.}~\bibnamefont {Bringel}}, \bibinfo {author} {\bibfnamefont
  {A.}~\bibnamefont {B{\"u}rger}}, \bibinfo {author} {\bibfnamefont
  {N.}~\bibnamefont {Nenoff}}, \bibinfo {author} {\bibfnamefont
  {G.}~\bibnamefont {Sch{\"o}nwa{\ss}er}}, \bibinfo {author} {\bibfnamefont
  {A.}~\bibnamefont {Singh}}, \bibinfo {author} {\bibfnamefont
  {D.}~\bibnamefont {Curien}}, \bibinfo {author} {\bibfnamefont
  {G.}~\bibnamefont {Hagemann}}, \bibinfo {author} {\bibfnamefont
  {B.}~\bibnamefont {Herskind}}, \bibinfo {author} {\bibfnamefont
  {G.}~\bibnamefont {Sletten}}, \bibinfo {author} {\bibfnamefont
  {P.}~\bibnamefont {Fallon}}, \bibinfo {author} {\bibfnamefont
  {A.}~\bibnamefont {G{\"o}rgen}}, \ and\ \bibinfo {author} {\bibfnamefont
  {P.}~\bibnamefont {Bednarczyk}},\ }\href
  {http://dx.doi.org/10.1140/epja/i2015-15050-y} {\bibfield  {journal}
  {\bibinfo  {journal} {Eur. Phys. J. A}\ }\textbf {\bibinfo {volume} {51}},\
  \bibinfo {eid} {50} (\bibinfo {year} {2015})}\BibitemShut {NoStop}%
\bibitem [{\citenamefont {Lee}(1990)}]{LEE1990c641}%
  \BibitemOpen
  \bibfield  {author} {\bibinfo {author} {\bibfnamefont {I.-Y.}\ \bibnamefont
  {Lee}},\ }\href {\doibase http://dx.doi.org/10.1016/0375-9474(90)91181-P}
  {\bibfield  {journal} {\bibinfo  {journal} {Nucl. Phys. A}\ }\textbf
  {\bibinfo {volume} {520}},\ \bibinfo {pages} {641c } (\bibinfo {year}
  {1990})}\BibitemShut {NoStop}%
\bibitem [{\citenamefont {Radford}(1995{\natexlab{a}})}]{rad1}%
  \BibitemOpen
  \bibfield  {author} {\bibinfo {author} {\bibfnamefont {D.}~\bibnamefont
  {Radford}},\ }\href {\doibase http://dx.doi.org/10.1016/0168-9002(95)00183-2}
  {\bibfield  {journal} {\bibinfo  {journal} {Nucl. Instrum. Meth. Phys. Res.
  A}\ }\textbf {\bibinfo {volume} {361}},\ \bibinfo {pages} {297 } (\bibinfo
  {year} {1995}{\natexlab{a}})}\BibitemShut {NoStop}%
\bibitem [{\citenamefont {Radford}(1995{\natexlab{b}})}]{rad2}%
  \BibitemOpen
  \bibfield  {author} {\bibinfo {author} {\bibfnamefont {D.}~\bibnamefont
  {Radford}},\ }\href {\doibase http://dx.doi.org/10.1016/0168-9002(95)00184-0}
  {\bibfield  {journal} {\bibinfo  {journal} {Nucl. Instrum. Meth. Phys. Res.
  A}\ }\textbf {\bibinfo {volume} {361}},\ \bibinfo {pages} {306 } (\bibinfo
  {year} {1995}{\natexlab{b}})}\BibitemShut {NoStop}%
\bibitem [{\citenamefont {Iacob}\ and\ \citenamefont
  {Duchene}(1997)}]{Iacob199757}%
  \BibitemOpen
  \bibfield  {author} {\bibinfo {author} {\bibfnamefont {V.}~\bibnamefont
  {Iacob}}\ and\ \bibinfo {author} {\bibfnamefont {G.}~\bibnamefont
  {Duchene}},\ }\href {\doibase 10.1016/S0168-9002(97)00872-3} {\bibfield
  {journal} {\bibinfo  {journal} {Nucl. Instrum. Meth. Phys. Res. A}\ }\textbf
  {\bibinfo {volume} {399}},\ \bibinfo {pages} {57 } (\bibinfo {year}
  {1997})}\BibitemShut {NoStop}%
\bibitem [{\citenamefont {Kr\"{a}mer-Flecken}\ \emph
  {et~al.}(1989)\citenamefont {Kr\"{a}mer-Flecken}, \citenamefont {Morek},
  \citenamefont {Lieder}, \citenamefont {Gast}, \citenamefont {Hebbinghaus},
  \citenamefont {J\"{a}ger},\ and\ \citenamefont
  {Urban}}]{KramerFlecken1989333}%
  \BibitemOpen
  \bibfield  {author} {\bibinfo {author} {\bibfnamefont {A.}~\bibnamefont
  {Kr\"{a}mer-Flecken}}, \bibinfo {author} {\bibfnamefont {T.}~\bibnamefont
  {Morek}}, \bibinfo {author} {\bibfnamefont {R.~M.}\ \bibnamefont {Lieder}},
  \bibinfo {author} {\bibfnamefont {W.}~\bibnamefont {Gast}}, \bibinfo {author}
  {\bibfnamefont {G.}~\bibnamefont {Hebbinghaus}}, \bibinfo {author}
  {\bibfnamefont {H.~M.}\ \bibnamefont {J\"{a}ger}}, \ and\ \bibinfo {author}
  {\bibfnamefont {W.}~\bibnamefont {Urban}},\ }\href {\doibase
  10.1016/0168-9002(89)90706-7} {\bibfield  {journal} {\bibinfo  {journal}
  {Nucl. Instrum. Meth. Phys. Res. A}\ }\textbf {\bibinfo {volume} {275}},\
  \bibinfo {pages} {333 } (\bibinfo {year} {1989})}\BibitemShut {NoStop}%
\bibitem [{\citenamefont {Chiara}\ \emph {et~al.}(2007)\citenamefont {Chiara},
  \citenamefont {Devlin}, \citenamefont {Ideguchi}, \citenamefont {LaFosse},
  \citenamefont {Lerma}, \citenamefont {Reviol}, \citenamefont {Ryu},
  \citenamefont {Sarantites}, \citenamefont {Pechenaya}, \citenamefont
  {Baktash}, \citenamefont {Galindo-Uribarri}, \citenamefont {Carpenter},
  \citenamefont {Janssens}, \citenamefont {Lauritsen}, \citenamefont {Lister},
  \citenamefont {Reiter}, \citenamefont {Seweryniak}, \citenamefont {Fallon},
  \citenamefont {G\"orgen}, \citenamefont {Macchiavelli}, \citenamefont
  {Rudolph}, \citenamefont {Stoitcheva},\ and\ \citenamefont
  {Ormand}}]{Chiara.75.054305}%
  \BibitemOpen
  \bibfield  {author} {\bibinfo {author} {\bibfnamefont {C.~J.}\ \bibnamefont
  {Chiara}}, \bibinfo {author} {\bibfnamefont {M.}~\bibnamefont {Devlin}},
  \bibinfo {author} {\bibfnamefont {E.}~\bibnamefont {Ideguchi}}, \bibinfo
  {author} {\bibfnamefont {D.~R.}\ \bibnamefont {LaFosse}}, \bibinfo {author}
  {\bibfnamefont {F.}~\bibnamefont {Lerma}}, \bibinfo {author} {\bibfnamefont
  {W.}~\bibnamefont {Reviol}}, \bibinfo {author} {\bibfnamefont {S.~K.}\
  \bibnamefont {Ryu}}, \bibinfo {author} {\bibfnamefont {D.~G.}\ \bibnamefont
  {Sarantites}}, \bibinfo {author} {\bibfnamefont {O.~L.}\ \bibnamefont
  {Pechenaya}}, \bibinfo {author} {\bibfnamefont {C.}~\bibnamefont {Baktash}},
  \bibinfo {author} {\bibfnamefont {A.}~\bibnamefont {Galindo-Uribarri}},
  \bibinfo {author} {\bibfnamefont {M.~P.}\ \bibnamefont {Carpenter}}, \bibinfo
  {author} {\bibfnamefont {R.~V.~F.}\ \bibnamefont {Janssens}}, \bibinfo
  {author} {\bibfnamefont {T.}~\bibnamefont {Lauritsen}}, \bibinfo {author}
  {\bibfnamefont {C.~J.}\ \bibnamefont {Lister}}, \bibinfo {author}
  {\bibfnamefont {P.}~\bibnamefont {Reiter}}, \bibinfo {author} {\bibfnamefont
  {D.}~\bibnamefont {Seweryniak}}, \bibinfo {author} {\bibfnamefont
  {P.}~\bibnamefont {Fallon}}, \bibinfo {author} {\bibfnamefont
  {A.}~\bibnamefont {G\"orgen}}, \bibinfo {author} {\bibfnamefont {A.~O.}\
  \bibnamefont {Macchiavelli}}, \bibinfo {author} {\bibfnamefont
  {D.}~\bibnamefont {Rudolph}}, \bibinfo {author} {\bibfnamefont
  {G.}~\bibnamefont {Stoitcheva}}, \ and\ \bibinfo {author} {\bibfnamefont
  {W.~E.}\ \bibnamefont {Ormand}},\ }\href {\doibase
  10.1103/PhysRevC.75.054305} {\bibfield  {journal} {\bibinfo  {journal} {Phys.
  Rev. C}\ }\textbf {\bibinfo {volume} {75}},\ \bibinfo {pages} {054305}
  (\bibinfo {year} {2007})}\BibitemShut {NoStop}%
\bibitem [{\citenamefont {Yoshinaga}\ and\ \citenamefont
  {Higashiyama}(2004)}]{PhysRevC.69.054309}%
  \BibitemOpen
  \bibfield  {author} {\bibinfo {author} {\bibfnamefont {N.}~\bibnamefont
  {Yoshinaga}}\ and\ \bibinfo {author} {\bibfnamefont {K.}~\bibnamefont
  {Higashiyama}},\ }\href {\doibase 10.1103/PhysRevC.69.054309} {\bibfield
  {journal} {\bibinfo  {journal} {Phys. Rev. C}\ }\textbf {\bibinfo {volume}
  {69}},\ \bibinfo {pages} {054309} (\bibinfo {year} {2004})}\BibitemShut
  {NoStop}%
\bibitem [{\citenamefont {Higashiyama}\ and\ \citenamefont
  {Yoshinaga}(2011)}]{PhysRevC.83.034321}%
  \BibitemOpen
  \bibfield  {author} {\bibinfo {author} {\bibfnamefont {K.}~\bibnamefont
  {Higashiyama}}\ and\ \bibinfo {author} {\bibfnamefont {N.}~\bibnamefont
  {Yoshinaga}},\ }\href {\doibase 10.1103/PhysRevC.83.034321} {\bibfield
  {journal} {\bibinfo  {journal} {Phys. Rev. C}\ }\textbf {\bibinfo {volume}
  {83}},\ \bibinfo {pages} {034321} (\bibinfo {year} {2011})}\BibitemShut
  {NoStop}%
\bibitem [{\citenamefont {Jia}\ \emph {et~al.}(2007)\citenamefont {Jia},
  \citenamefont {Zhang},\ and\ \citenamefont {Zhao}}]{PhysRevC.76.054305}%
  \BibitemOpen
  \bibfield  {author} {\bibinfo {author} {\bibfnamefont {L.~Y.}\ \bibnamefont
  {Jia}}, \bibinfo {author} {\bibfnamefont {H.}~\bibnamefont {Zhang}}, \ and\
  \bibinfo {author} {\bibfnamefont {Y.~M.}\ \bibnamefont {Zhao}},\ }\href
  {\doibase 10.1103/PhysRevC.76.054305} {\bibfield  {journal} {\bibinfo
  {journal} {Phys. Rev. C}\ }\textbf {\bibinfo {volume} {76}},\ \bibinfo
  {pages} {054305} (\bibinfo {year} {2007})}\BibitemShut {NoStop}%
\bibitem [{\citenamefont {Petrache}\ \emph {et~al.}(1996)\citenamefont
  {Petrache}, \citenamefont {Sun}, \citenamefont {Bazzacco}, \citenamefont
  {Lunardi}, \citenamefont {Alvarez}, \citenamefont {Venturelli}, \citenamefont
  {De~Acu\~na}, \citenamefont {Maron}, \citenamefont {Rao}, \citenamefont
  {Podoly\'ak},\ and\ \citenamefont {Oliveira}}]{pet1996}%
  \BibitemOpen
  \bibfield  {author} {\bibinfo {author} {\bibfnamefont {C.~M.}\ \bibnamefont
  {Petrache}}, \bibinfo {author} {\bibfnamefont {Y.}~\bibnamefont {Sun}},
  \bibinfo {author} {\bibfnamefont {D.}~\bibnamefont {Bazzacco}}, \bibinfo
  {author} {\bibfnamefont {S.}~\bibnamefont {Lunardi}}, \bibinfo {author}
  {\bibfnamefont {C.~R.}\ \bibnamefont {Alvarez}}, \bibinfo {author}
  {\bibfnamefont {R.}~\bibnamefont {Venturelli}}, \bibinfo {author}
  {\bibfnamefont {D.}~\bibnamefont {De~Acu\~na}}, \bibinfo {author}
  {\bibfnamefont {G.}~\bibnamefont {Maron}}, \bibinfo {author} {\bibfnamefont
  {M.~N.}\ \bibnamefont {Rao}}, \bibinfo {author} {\bibfnamefont
  {Z.}~\bibnamefont {Podoly\'ak}}, \ and\ \bibinfo {author} {\bibfnamefont
  {J.~R.~B.}\ \bibnamefont {Oliveira}},\ }\href {\doibase
  10.1103/PhysRevC.53.R2581} {\bibfield  {journal} {\bibinfo  {journal} {Phys.
  Rev. C}\ }\textbf {\bibinfo {volume} {53}},\ \bibinfo {pages} {R2581}
  (\bibinfo {year} {1996})}\BibitemShut {NoStop}%
\bibitem [{\citenamefont {Petrache}\ and\ \citenamefont {{et
  al.}}(2015)}]{pet2015}%
  \BibitemOpen
  \bibfield  {author} {\bibinfo {author} {\bibfnamefont {C.~M.}\ \bibnamefont
  {Petrache}}\ and\ \bibinfo {author} {\bibnamefont {{et al.}}},\ }\href@noop
  {} {}\bibinfo {howpublished} {Unpublished} (\bibinfo {year}
  {2015})\BibitemShut {NoStop}%
\bibitem [{\citenamefont {Zhao}\ \emph {et~al.}(2010)\citenamefont {Zhao},
  \citenamefont {Li}, \citenamefont {Yao},\ and\ \citenamefont
  {Meng}}]{Zhao2010Phys.Rev.C54319}%
  \BibitemOpen
  \bibfield  {author} {\bibinfo {author} {\bibfnamefont {P.~W.}\ \bibnamefont
  {Zhao}}, \bibinfo {author} {\bibfnamefont {Z.~P.}\ \bibnamefont {Li}},
  \bibinfo {author} {\bibfnamefont {J.~M.}\ \bibnamefont {Yao}}, \ and\
  \bibinfo {author} {\bibfnamefont {J.}~\bibnamefont {Meng}},\ }\href {\doibase
  10.1103/PhysRevC.82.054319} {\bibfield  {journal} {\bibinfo  {journal} {Phys.
  Rev. C}\ }\textbf {\bibinfo {volume} {82}},\ \bibinfo {pages} {054319}
  (\bibinfo {year} {2010})}\BibitemShut {NoStop}%
\bibitem [{\citenamefont {Zhao}\ \emph {et~al.}(2015)\citenamefont {Zhao},
  \citenamefont {Zhang},\ and\ \citenamefont
  {Meng}}]{Zhao2015PhysRevC.92.034319}%
  \BibitemOpen
  \bibfield  {author} {\bibinfo {author} {\bibfnamefont {P.~W.}\ \bibnamefont
  {Zhao}}, \bibinfo {author} {\bibfnamefont {S.~Q.}\ \bibnamefont {Zhang}}, \
  and\ \bibinfo {author} {\bibfnamefont {J.}~\bibnamefont {Meng}},\ }\href
  {\doibase 10.1103/PhysRevC.92.034319} {\bibfield  {journal} {\bibinfo
  {journal} {Phys. Rev. C}\ }\textbf {\bibinfo {volume} {92}},\ \bibinfo
  {pages} {034319} (\bibinfo {year} {2015})}\BibitemShut {NoStop}%
\end{thebibliography}%

\end{document}